\documentclass[english,notitlepage,nofootinbib,reprint]{revtex4-1}
\usepackage[T1]{fontenc}
\usepackage[latin9]{inputenc}
\usepackage{color}
\usepackage{amsmath}
\usepackage{graphicx}
\usepackage{esint}
\usepackage{bm}

\makeatletter

\providecommand{\tabularnewline}{\\}

\usepackage{slashed}
\usepackage{cases}
\usepackage[colorlinks=true,citecolor=blue,linkcolor=blue]{hyperref}

\makeatother

\newcommand\vecl[1]{{\bf{#1}}}         		
\newcommand\vecs[1]{\boldsymbol{#1}}

\newcommand\vSIbb[0]{$2\nu_\text{SI}\beta\beta$}

\newcommand\suppmat[0]{Appendix}

\usepackage{babel}

\begin{document}

\title{Neutrino Self-Interactions and Double Beta Decay}

\author{Frank F.\ Deppisch$^{a}$, Lukas Graf$^{b}$, Werner Rodejohann$^{b}$, Xun-Jie Xu$^{b}$}

\affiliation{\textcolor{black}{$^{a}$Department of Physics and Astronomy, University College London,
Gower Street, London WC1E 6BT, UK}}

\affiliation{\textcolor{black}{$^{b}$Max-Planck-Institut f\"ur Kernphysik, Postfach 103980, D-69029, Heidelberg, Germany}}

\begin{abstract}
\noindent
Neutrino Self-Interactions ($\nu$SI) beyond the Standard Model are an attractive possibility to soften cosmological constraints on neutrino properties and also to explain the tension in late and early time measurements of the Hubble expansion rate. The required strength of $\nu$SI to explain the $4\sigma$ Hubble tension is in terms of a point-like effective four-fermion coupling that can be as high as $10^9\, G_F$, where $G_F$ is the Fermi constant. In this work, we show that such strong $\nu$SI can cause significant effects in two-neutrino double beta decay, leading to an observable enhancement of decay rates and to spectrum distortions. We analyze self-interactions via an effective operator as well as when mediated by a light scalar. Data from observed two-neutrino double beta decay is used to constrain $\nu$SI, which rules out the regime around $10^9\, G_F$.

\end{abstract}
\maketitle

\section{Introduction}
\noindent
The discrepancy between Cosmic Microwave Background (CMB) and local measurements of the Hubble constant, known as the Hubble tension, has grown to about $4\sigma$ \cite{Riess:2016jrr, Shanks:2018rka, Riess:2018kzi, Aghanim:2018eyx, Riess:2019cxk}. If indeed a physical fact, it would imply that non-standard particle physics or cosmology is required. Introducing a neutrino self-interaction ($\nu$SI), 
i.e.\ a four-neutrino contact interaction, to inhibit neutrino free-streaming in the early Universe can resolve the Hubble tension. The required strength of $\nu$SI needs to be much larger than the Fermi effective interactions predicted in the Standard Model (SM) \cite{Cyr-Racine:2013jua, Lancaster:2017ksf, Oldengott:2017fhy, Kreisch:2019yzn, Park:2019ibn}. Writing the interaction as $G_S \left(\nu\nu\right)\left(\nu\nu\right)$,\footnote{Here we adopt the Weyl spinor notation, with $\nu$ being a two-component spinor and the combination $(\nu\nu)$ is a scalar product.} there are two regimes for the coupling $G_S$: a strongly interacting regime with $G_S = 3.83_{-0.54}^{+1.22}\times 10^9\, G_F$ and a moderately interacting regime with $1.3\times 10^6 < G_S/G_F < 1.1\times 10^8$~\cite{Kreisch:2019yzn}.

The required new interaction is thus clearly a strong one. If taken seriously, it would indicate the presence of New Physics at a scale $G_S^{-1/2} \sim 10$~MeV~--~$1$~GeV. Such strong $\nu$SI have drawn considerable attention \cite{Hasenkamp:2016pme, Huang:2017egl, Bakhti:2018avv, Blinov:2019gcj, deGouvea:2019phk, Das:2017iuj, Dighe:2017sur, Ko:2019ssx, Shalgar:2019rqe, Forastieri:2019cuf,Lyu:2020lps}, but in general they are difficult to probe in laboratory experiments due to the absence of electrons or quarks involved. Assuming that $\nu$SI are mediated by new light bosons, existing constraints come from Big Bang Nucleosynthesis \cite{Boehm:2012gr, Kamada:2015era, Huang:2017egl}, pion/kaon decay \cite{Barger:1981vd, Lessa:2007up, Pasquini:2015fjv}, $Z$ invisible decay \cite{Berryman:2018ogk, Brdar:2020nbj}, and supernova neutrinos \cite{Das:2017iuj, Dighe:2017sur, Ko:2019ssx, Shalgar:2019rqe}. There is currently no direct constraint on the $\nu$SI operator without any assumption on its origin.

\begin{figure}[t]
	\centering
	\includegraphics[width=0.85\columnwidth]{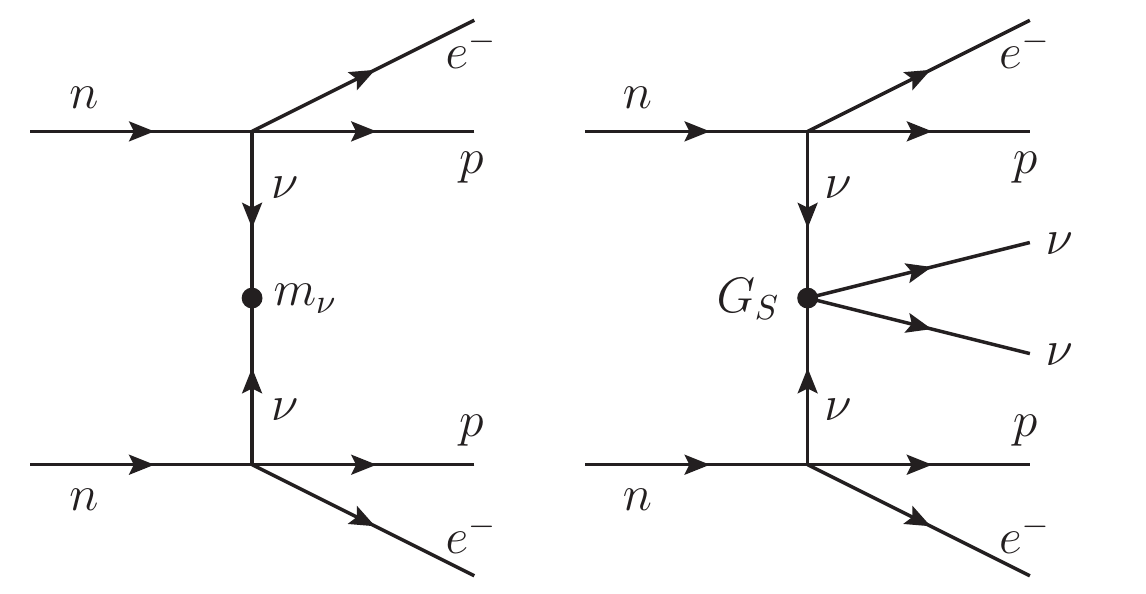}
	\caption{Left: Neutrinoless double beta decay via Majorana neutrino exchange. Right: $\nu$SI-induced double beta decay.}
	\label{fig:feyn}
\end{figure}

In this work, we propose to search for $\nu$SI in double beta decay experiments\footnote{For future prospects in beta decay experiments, see Ref.\ \cite{Arcadi:2018xdd}.}. These experiments search for the lepton number violating, and thus SM-forbidden, neutrinoless double beta decay ($0\nu\beta\beta$) \cite{Deppisch:2012nb, Dolinski:2019nrj}. 
The standard diagram of this process is the exchange of a massive Majorana neutrino, see Fig.~\ref{fig:feyn}~(left). As part of this effort, the SM-allowed two-neutrino double beta ($2\nu\beta\beta$) decay is measured with increasing precision and may itself be used to probe physics beyond the Standard  Model \cite{Deppisch:2020mxv}. 
In the presence of $\nu$SI, two neutrinos can be emitted via the effective $\nu$SI operator, see Fig.~\ref{fig:feyn}~(right). The final state of this $\nu$SI-induced double beta (\vSIbb) decay is identical to that of $2\nu\beta\beta$ decay. 
We will here discuss the total decay rate of \vSIbb{} decay and the energy as well as angular distributions of the emitted electrons. Such a study is warranted because a dimensional analysis estimate reveals that the total decay rates of $2\nu\beta\beta$ and \vSIbb{} are $\Gamma_{2\nu} \sim G_F^4 (0.1p_F)^{-2} Q^{11}$ and $\Gamma_{\nu\text{SI}} \sim G_S^2 G_F^4 p_F^2 Q^{11}$, respectively. Here, the Fermi momentum $p_F \approx 100$~MeV represents the nuclear scale and $Q \approx (1-4)$~MeV is the isotope-dependent kinetic energy release ($Q$-value) in double beta decay. 
The factor 0.1 in $\Gamma_{2\nu}$ takes into account that without a virtual neutrino line only states up to 10 MeV are excited in $2\nu\beta\beta$ decay. Assuming that \vSIbb{} rates of order $\Gamma_{\nu\text{SI}} \gtrsim \Gamma_{2\nu}$ can be seen experimentally, it is expected that couplings $G_S \gtrsim 10 \, p_F^{-2} \approx 10^8\, G_F$ can be probed.

In previous studies of double beta decay, indirect constraints on $\nu$SI mediated by light scalars were obtained \cite{Burgess:1992dt, Burgess:1993xh, Gando:2012pj, Agostini:2015nwa, Blum:2018ljv, Cepedello:2018zvr, Brune:2018sab, Farzan:2018gtr}. It was assumed that the scalar is emitted in the decay, hence is lighter than the $Q$-value of double beta decay. For a scalar particle $\phi$ that couples with strength $g_\phi$ to two electron neutrinos, one finds from searches for so-called Majoron emitting double beta decays that $g_\phi \lesssim 10^{-4} - 10^{-5}$ \cite{Gando:2012pj, Agostini:2015nwa}. Taking $m_{\phi} = 1$~MeV, this bound on the Yukawa coupling corresponds to $G_S \lesssim (10 - 10^3)\, G_F$. If $\nu$SI are not mediated by light scalars or the scalar mass is larger than the $Q$-value, this bound does not apply. In this case, the effect of $\nu$SI operators on double beta decay becomes more important, which we will investigate here. 

In the next section we will study \vSIbb{} decay in the effective operator language. In Sec.~\ref{sec:dist} we will generate the operator with an $s$-channel mediator whose mass is larger than the $Q$-value, and show how the distributions are affected. We conclude in Sec.~\ref{sec:concl}, and various technical details are delegated to the \suppmat.

\begin{table*}
\caption{\label{tab:t}Estimate of \vSIbb{} decay rates for several isotopes. Here, $Q$ is the corresponding $Q$-value, $T_{1/2}^{2\nu}$ represents the experimental $2\nu\beta\beta$ decay half-lives adopted from Ref.~\cite{Barabash:2019nnr} that can be translated to the experimental $2\nu\beta\beta$ decay rates using $\Gamma_{2\nu}^{{\rm ex}}=\log 2/T_{1/2}^{2\nu}$ and $\Gamma_{\nu\text{SI}}$ denotes the theoretical prediction for the \vSIbb{} decay rates computed from Eq.~\eqref{eq:gamma_vSI}, assuming $G_S = 3.83\times 10^9\, G_F$. Bounds on $G_S$ obtained according to Eq.~\eqref{eq:compare} are presented in the last row. Nuclear matrix element values from IBM-2~\cite{Barea:2015kwa} are used to obtain the values in this table. 
}
\begin{ruledtabular}
\begin{tabular}{ccccccc}
	& $^{48}{\rm Ca}$  & $^{76}{\rm Ge}$  & $^{136}{\rm Xe}$  & $^{100}{\rm Mo}$  & $^{128}{\rm Te}$  & $^{130}{\rm Te}$\tabularnewline
	\hline 
	$Q/$MeV & 4.263 \cite{Redshaw:2012jp} & 2.039 \cite{Rahaman:2007ng} & 2.458 \cite{Redshaw:2007un} & 3.034 \cite{Rahaman:2007ng} & 0.8659 \cite{Scielzo:2009nh} & 2.527 \cite{Rahaman:2011zz}\tabularnewline
	$T^{2\nu}_{1/2}$/year & $5.30\times10^{19}$ & $1.88\times10^{21}$ & $2.17\times10^{21}$ & $6.88\times10^{18}$ & $2.25\times10^{24}$ & $7.91\times10^{20}$\tabularnewline
	$\left(\Gamma_{\nu{\rm SI}}\right)^{-1}$/year & $2.52\times10^{18}$ & $1.42\times10^{20}$ & $1.55\times10^{19}$ & $2.94\times10^{18}$ & $4.04\times10^{22}$ & $9.08\times10^{18}$ \tabularnewline
	$\Gamma_{\nu{\rm SI}}/\Gamma_{2\nu}^{{\rm ex}}$ & $30.3$ & $19.0$ & $203$ & $3.38$ & $80.4$ & $126$\tabularnewline
	$G_S/G_F<$ & $0.84\times 10^9$ & $1.05\times 10^9$ & $0.32\times 10^9$ & $2.50\times 10^9$ & $0.51\times 10^9$ & $0.41\times 10^9$ \tabularnewline
	\end{tabular}
\end{ruledtabular}
\end{table*}
\section{$\nu$SI-induced double beta decay \label{sec:basic}}
\noindent
In the standard $0\nu\beta\beta$ mechanism, two neutrinos produced in double beta decay annihilate due to a Majorana mass term, leaving only electrons in the leptonic final states, as shown in Fig.~\ref{fig:feyn}~(left). Under the presence of $\nu$SI operators,
\begin{align}
	{\cal L}_{\nu\text{SI}}^\text{LNC} &= G_S(\nu_e\nu_e)(\overline{\nu}_\alpha\overline{\nu}_\beta), \quad \text{or} \nonumber\\
	{\cal L}_{\nu\text{SI}}^\text{LNV} &= G_S(\nu_e\nu_e)(\nu_\alpha\nu_\beta),
\label{eq:L}
\end{align}
where $\alpha,\beta = e,\mu,\tau$ are flavor indices, Fig.~\ref{fig:feyn}~(right) implies that the two electron-antineutrinos ($\bar\nu_e$) generated by neutron decay can take part in the $\nu$SI interaction and convert in the \vSIbb{} process to $\nu_\alpha \nu_\beta$ or $\overline{\nu}_\alpha \overline{\nu}_\beta$. Note that both the lepton number conserving (LNC) and violating (LNV) interactions in Eq.~\eqref{eq:L} can lead to \vSIbb.

Assuming that the momenta of leptonic final states are negligible compared to the momenta of the neutrino propagators (the typical values of the former and the latter are of order $Q = {\cal O}(1)$~MeV and $p_F =  {\cal O}(100)$~MeV, respectively), it can be shown that the two processes in Fig.~\ref{fig:feyn} share the same nuclear matrix elements (NMEs), see  \suppmat~\ref{sec:appA}. Consequently, we can compute the decay rate of \vSIbb{} using the NME of $0\nu\beta\beta$: 
\begin{equation}
	\Gamma_{\nu{\rm SI}} = \left|\frac{G_S m_e}{2R}\right|^2 \mathcal{G}_{\nu\mathrm{SI}}|\mathcal{M}_{0\nu}|^2 .
\label{eq:gamma_vSI}
\end{equation}
Here $m_e$ denotes the electron mass and $R = 1.2 A^{1/3}$~fm is the radius of the nucleus with nucleon number $A$. The structure of the $0\nu\beta\beta$ NME $\mathcal{M}_{0\nu}$ is explained in \suppmat~\ref{sec:appA-nuclear}. The quantity ${\cal G}_{\nu\text{SI}}$ is the \vSIbb{} phase space factor, which is derived in \suppmat~\ref{sec:appA-leptonic}. It  reads
\begin{equation}
	\mathcal{G}_{\nu{\rm SI}} = \frac{2c_{\nu\text{SI}}}{15}\int\! dp_1 dp_2
	p_1^2p_2^2(Q-T_{12})^5 F^2(p_1,p_2),
\label{eq:IvSI}
\end{equation}
where $p_1$ and $p_2$ are the momenta of the two electrons.
Neglecting the final state lepton momenta in the calculation of the $2\nu\beta\beta$ and \vSIbb{} NMEs, the phase space factors are related as ${\cal G}_{\nu\text{SI}}={\cal G}_{2\nu}/(4\pi)^2$.
The $Q$-value is given in Tab.~\ref{tab:t} for various isotopes and $F^2(p_1,p_2)$ stands for the Fermi function correction caused by the Coulomb potential of the nucleus. Finally, $T_{12} = E_1 + E_2 - 2m_e$ is the total kinetic energy of both electrons, implicitly depending on $p_1$ and $p_2$, and neutrino masses in the final state have been neglected. The constant $c_{\nu\text{SI}}$ appearing in the above equation reads
\begin{equation}
	c_{\nu\text{SI}} = \frac{G_F^4\cos^4\!\theta_C}{256\pi^9m_e^2},
\end{equation}
where $\theta_C$ denotes the Cabibbo angle. 
Note that the electron mass $m_e$ and nuclear radius $R$ are included in Eq.~\eqref{eq:gamma_vSI} so that the normalization of the NME and phase space factor conforms with that adopted in the literature.


Using Eqs.~\eqref{eq:gamma_vSI} and \eqref{eq:IvSI}, it is straightforward to compute $\Gamma_{\nu{\rm SI}}$. It should be noted, however, that the electron spectrum of \vSIbb{} is very similar to that of $2\nu\beta\beta$ decay.
We will comment below on the potential differences arising in the case of a light $s$-channel scalar mediator inducing the $\nu$SI. 

If the energy and angular resolution of detectors cannot distinguish the electron spectrum of \vSIbb{} decay from that of $2\nu\beta\beta$ decay, then only the change of the total decay rate can be probed.
The total rate $\Gamma_{2\nu}$ of $2\nu\beta\beta$ decay has been measured precisely for many isotopes. For example, the $2\nu\beta\beta$ rate of $^{136}$Xe has been measured to a $3\%$ level \cite{Albert:2013gpz}. Nonetheless, there remains a considerable uncertainty in the theoretical prediction of the $2\nu\beta\beta$ decay rate arising from the NMEs. Writing the theoretical prediction for the total decay rate approximately as
\begin{equation}
	\Gamma_{2\nu} + \Gamma_{\nu\text{SI}} \approx 
	\left(|\mathcal{M}_{2\nu}|^2 
	+ \left|\frac{G_S m_e}{2R}\right|^2 \frac{|\mathcal{M}_{0\nu}|^2}{4\pi^2}\right)\mathcal{G}_{2\nu},
\label{eq:summed-rate}
\end{equation}
the sensitivity to $G_S$ will largely depend on the uncertainty of the NME ratio $|\mathcal{M}_{0\nu}|^2 / |\mathcal{M}_{2\nu}|^2$. While some of the nuclear uncertainties are expected to drop out from this, unresolved issues such as the quenching of the effective nuclear axial coupling $g_A$ in $0\nu\beta\beta$ decay likely provide a major error source. Note that in the above equation, we neglect the effect of interference between the $2\nu\beta\beta$ and \vSIbb{} diagrams. If two electron anti-neutrinos are being emitted in \vSIbb{} decay, such an interference will generally take place.

We proceed by constraining the new interaction requiring that the \vSIbb{} rate is less than the one for $2\nu\beta\beta$,
\begin{equation}
	\Gamma_{\nu{\rm SI}}/\Gamma_{2\nu}^{{\rm ex}} < 1,
\label{eq:compare}
\end{equation}
where $\Gamma_{2\nu}^{{\rm ex}}$ stands for the experimentally measured value of $2\nu\beta\beta$. 
This roughly corresponds to an assumed uncertainty in the NME ratio $|\mathcal{M}_{0\nu}|^2 / |\mathcal{M}_{2\nu}|^2$ within a factor of two. If this uncertainty can be reduced in future theory NME determinations, the sensitivity on $G_S$ will improve accordingly. As mentioned, the uncertainty depends on $g_A$. In our calculations we implicitly assume the unquenched value $g_A = 1.269$ as used in the calculation of the NMEs.

\begin{figure}
	\centering
	\includegraphics[width=\columnwidth]{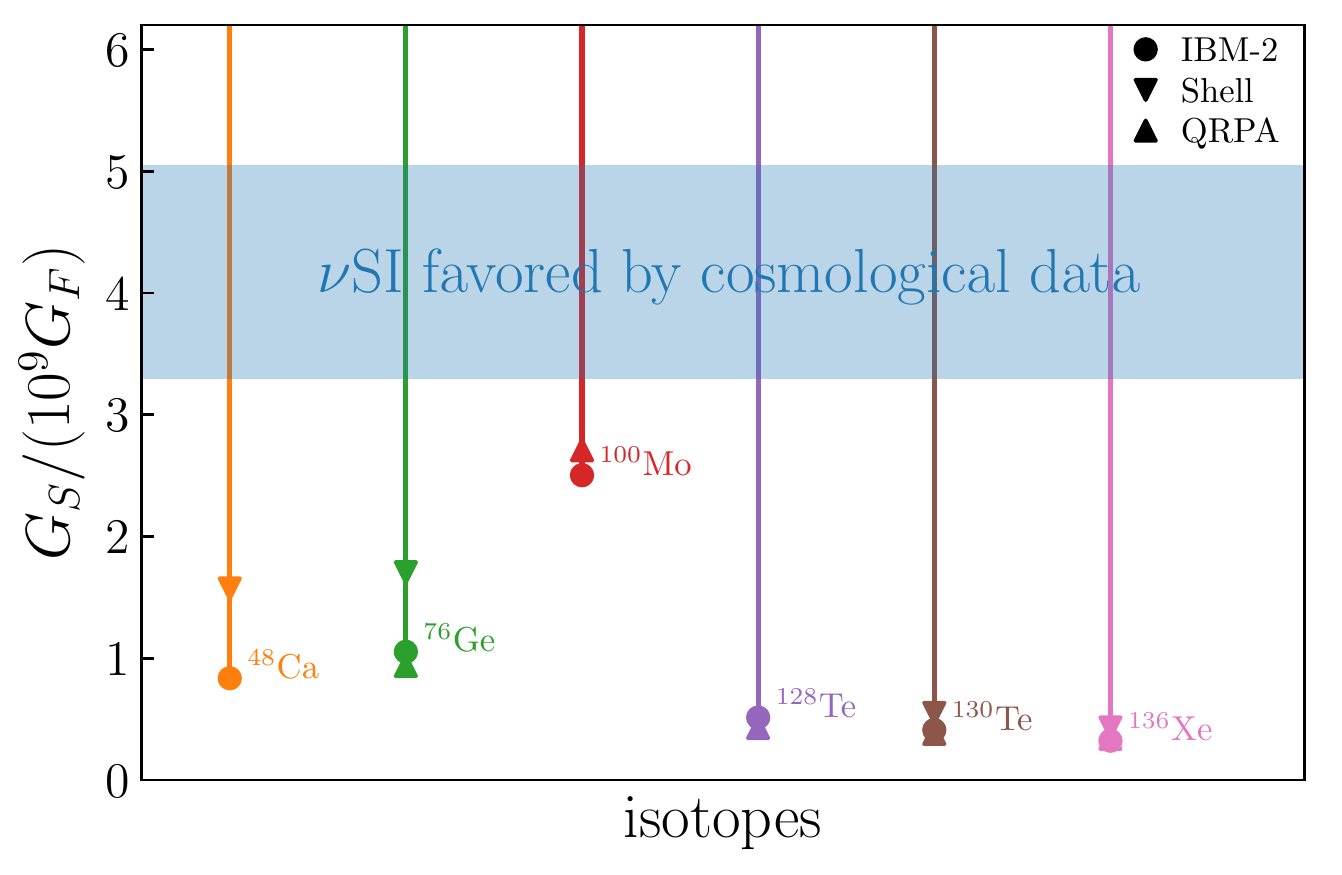}
	\caption{\label{fig:limit} Upper limit on the $\nu$SI coupling $G_S$ from $2\nu\beta\beta$ decay data for several isotopes and three different NME calculations as indicated. The blue band corresponds to the strongly interacting regime $G_S = 3.83_{-0.54}^{+1.22}\times 10^9\, G_F$ favored by cosmological data, which here is excluded by observations of $2\nu\beta\beta$ of various isotopes.}
\end{figure}

Taking the best-fit value of $G_S = 3.83\times 10^{9}\,G_F$ of the strongly interacting regime and the Interacting Boson Model (IBM-2) NMEs \cite{Barea:2015kwa}, we compute the decay rate $\Gamma_{\nu{\rm SI}}$ and compare it with $\Gamma_{2\nu}^{{\rm ex}}$ in Tab.~\ref{tab:t}. By requiring $\Gamma_{\nu{\rm SI}}/\Gamma_{2\nu}^{{\rm ex}} < 1$, we obtain the corresponding constraints on $G_S$, which is presented in Fig.~\ref{fig:limit}. Here, we also use NME values computed in the Interacting Shell Model (Shell) \cite{Menendez:2017fdf} and  Quasi-particle Random Phase Approximation (QRPA) model \cite{Hyvarinen:2015bda}; the corresponding limits on $G_S$ are shown in Fig.~\ref{fig:limit}, indicating the uncertainty arising from nuclear theory uncertainties. As one can see, the strongly interacting regime for $G_S$ favored by the cosmological data causes $\Gamma_{\nu{\rm SI}}/\Gamma_{2\nu}^{{\rm ex}} > 1$ for all the isotopes listed in Tab.~\ref{tab:t}. For some isotopes, $\Gamma_{\nu{\rm SI}}$ can be even one or two orders of magnitude higher than $\Gamma_{2\nu}^{{\rm ex}}$. Even including the theoretical NME uncertainties, most isotopes can fully exclude the cosmologically favored strongly interacting regime band, given the premise that two $\nu_e$ are involved in the $\nu$SI.

\section{Energy and angular distributions}
\label{sec:dist}

\begin{figure}[t]
	\centering
	\includegraphics[width=0.9\columnwidth]{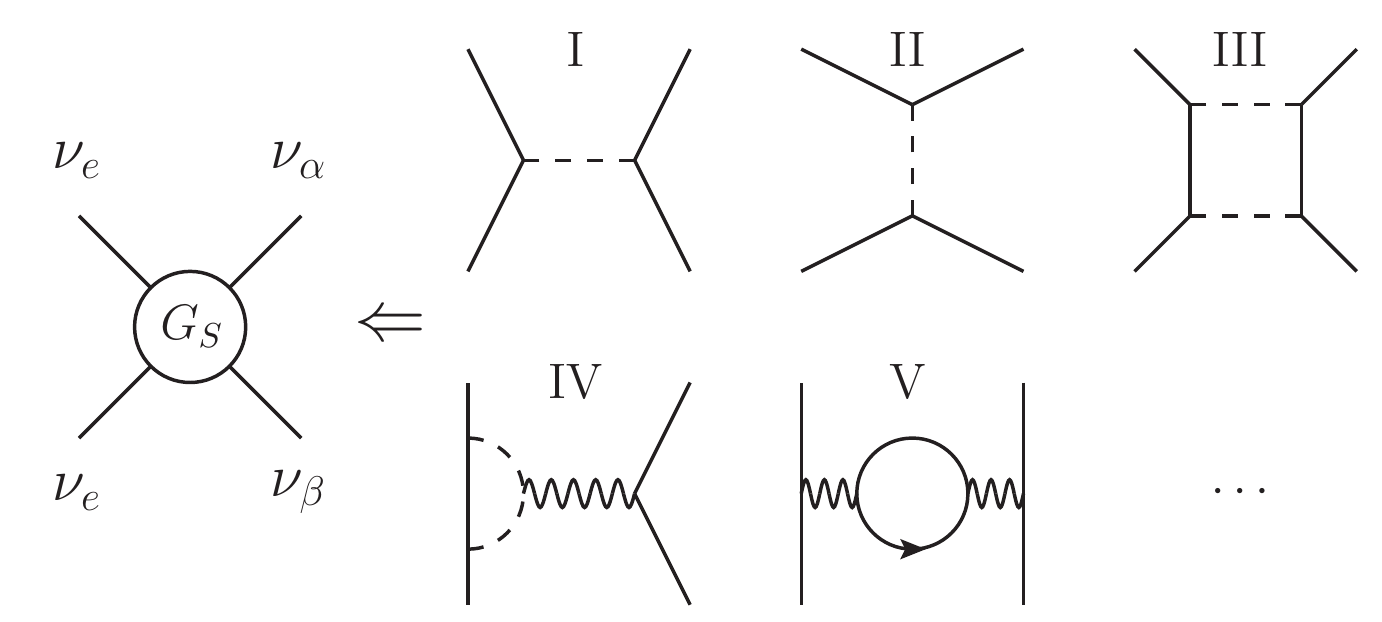}
	\caption{\label{fig:UV_complete}$\nu$SI generation at tree and loop level. Here, scalar (vector) lines may also be replaced by vector (scalar) lines.}
\end{figure}

\noindent
We now consider possible distortions of the electron energy and angular distributions arising from the $\nu$SI-induced contribution. For an exact contact interaction of four neutrinos and neglecting final state lepton momenta, one can show that the electron spectra of \vSIbb{} decay are the same as that of $2\nu\beta\beta$ decay, see \suppmat~\ref{sec:appA-leptonic}. However, considering that the $\nu$SI operator may be generated by light mediators, the corresponding energy dependence of $G_S$, can cause observable spectral distortions, as we shall discuss below. We should mention here that the spectral distortions depend on the underlying model for  $\nu$SI, which currently still lacks comprehensive exploration. There are various possibilities to generate the $\nu$SI effective operator, as shown in Fig.~\ref{fig:UV_complete}, where both tree and one-loop level diagrams are illustrated. 

At tree level, the $\nu$SI operator can be opened via either an $s$-channel (diagram I) or a $t$-channel (diagram II) scalar mediator. For vector mediators, most of the discussions below apply as well\footnote{However, if the $s$-channel mediator is a vector boson, the process would be suppressed by the tiny neutrino masses due to the required chirality flipping \cite{Carone:1993jv}.}. For the $s$- and $t$-channel diagrams, $G_S$ has the following energy dependence
\begin{numcases}{G_S =}    
	\frac{-m_\phi^2}{s - m_\phi^2} G_S^0 \quad (s\text{-channel}), 
\label{eq:st-s}    \\    
	\frac{ m_\phi^2}{t + m_\phi^2} G_S^0 \quad (t\text{-channel}),
\label{eq:st-t} 
\end{numcases}
where $m_\phi$ is the mediator mass and $s\equiv p^2$, $t\equiv -q^2$ with $p$ and $q$ being the momenta of the mediators in the tree diagrams. In the context of \vSIbb, they are of order $t \sim p_F^2$ and $s \lesssim Q^2$, respectively. The values of $G_S$ at zero momentum transfer are denoted as $G_S^0 = g^2_\phi/m^2_\phi$, with the coupling $g_\phi$ between $\phi$ and the neutrinos. Note that in Eq.~\eqref{eq:st-s} we omit the small effect of the $\phi$ decay width.

At the one-loop level, the $\nu$SI operators can be generated e.g. by the box diagram in Fig.~\ref{fig:UV_complete}. The corresponding energy dependence of $G_S$ is much more complicated than in the tree-level case. In general, it depends on both $s = p^2$ and $t = -q^2$. 
However, for most loop diagrams, there are no simple analytical expressions similar to Eqs.~\eqref{eq:st-s} and \eqref{eq:st-t}. For the box diagram, we can obtain a simple result assuming all the particles running in the loop have the same mass $m_\phi$ and that $s \ll t \ll m_\phi^2$. With these assumptions, following the calculation in Ref.~\cite{Bischer:2018zbd}, we get
\begin{equation}
	G_S = G_S^0\left(1 - \frac{3}{10}\frac{t}{m_\phi^2} + \cdots\right).
\label{eq:box-1}
\end{equation}
Compared to Eq.~\eqref{eq:st-t}, where the expansion in $t$ yields $G_S = G_S^0(1 - t/m_\phi^2 + \cdots)$, the $t/m_\phi^2$ term in Eq.~\eqref{eq:box-1} has a different coefficient but the same sign. In addition to the box diagram, other one-loop diagrams are also possible, as illustrated by diagrams IV and V in Fig.~\ref{fig:UV_complete}. Such diagrams can be roughly regarded as tree-level diagrams with energy-dependent couplings or mediator masses, which may cause more elusive effects in probing $\nu$SI in experiments of different energy scales. Here we only mention these possibilities and refrain from further discussions\footnote{We note that $\nu$SI may also lead to significant corrections to the neutrino self-energy, which is not fully identical to the neutrino mass in $0\nu\beta\beta$~\cite{Rodejohann:2019quz}. The effect is quite model-dependent, and can be studied if a complete model of $\nu$SI has been constructed.}.

Among the aforementioned possibilities, only the $s$-channel case in Eq.~\eqref{eq:st-s} can be analyzed without involving novel nuclear physics calculations.
Other $t$-dependent scenarios necessarily involve integrals over $q$ that are different from the one in $0\nu\beta\beta$ decay, which calls for a dedicated study in the future. 
Here we proceed only with the $s$-channel case, specifically for $m_\phi \gtrsim Q$. While the $t$-channel may also contribute in this regime, its effect is expected to be considerably smaller due to the large $t$ suppression in the propagator.
For $G_S$ in Eq.~\eqref{eq:st-s}, we have derived the \vSIbb{} differential decay rate in \suppmat~\ref{sec:appA-leptonic} yielding the dependence
\begin{align}
	\frac{d\Gamma_{\nu{\rm SI}}}{dp_1 dp_2 d\!\cos\theta_{12}}
	&\propto |G_S^0|^2 p_1^2 p_2^2 F^2(p_1,p_2) \nonumber\\ 
	&\times I_s(T_{12}) \left(1 - \beta_1\beta_2\cos\theta_{12}\right).
\label{eq:s-5-1}
\end{align}
Here, $\cos\theta_{12} = {\bf p}_1\cdot {\bf p}_2/(p_1 p_2)$ with the angle $0 \leq \theta_{12} \leq \pi$ between the two emitted electrons and $\beta_i = p_i/E_i$ are the electron velocities. The effect of the $s$-channel mediating scalar is captured by the function
\begin{equation}
	I_s(T_{12}) = \frac{Q - T_{12}}{4(2\pi)^4} \left(\xi\frac{2 + \cos\xi}{\sin\xi} - 3\right),
\label{eq:Is}
\end{equation}
where $\xi = 2\arcsin((Q - T_{12})/m_\phi)$. It is a function of the total electron kinetic energy $T_{12}$,  and as we will see it can cause distortions of both the energy and angular distributions of the electrons. 
In the limit $m_\phi \to \infty$, the effective operator is recovered and the dependence approaches $I_s(T_{12}) \propto (Q - T_{12})^5$ yielding a phase space factor equivalent to $2\nu\beta\beta$ decay.

We first consider the energy distribution. All modern double beta decay experiments measure the differential decay rate $d\Gamma_{\nu\text{SI}}/dT$ with respect to the total electron kinetic energy. This rate is computed by integrating over $\cos\theta_{12}$, $p_1$ and $p_2$ with the total kinetic energy $T = T_{12}$ fixed at a given value.
%
As noted, in the limit $m_\phi\to\infty$, $d\Gamma_{\nu\text{SI}}/dT$ will have the same energy distribution as that of $2\nu\beta\beta$ decay. 
%
%

%
\begin{figure}[t]
	\centering	
	\includegraphics[width=0.95\columnwidth]{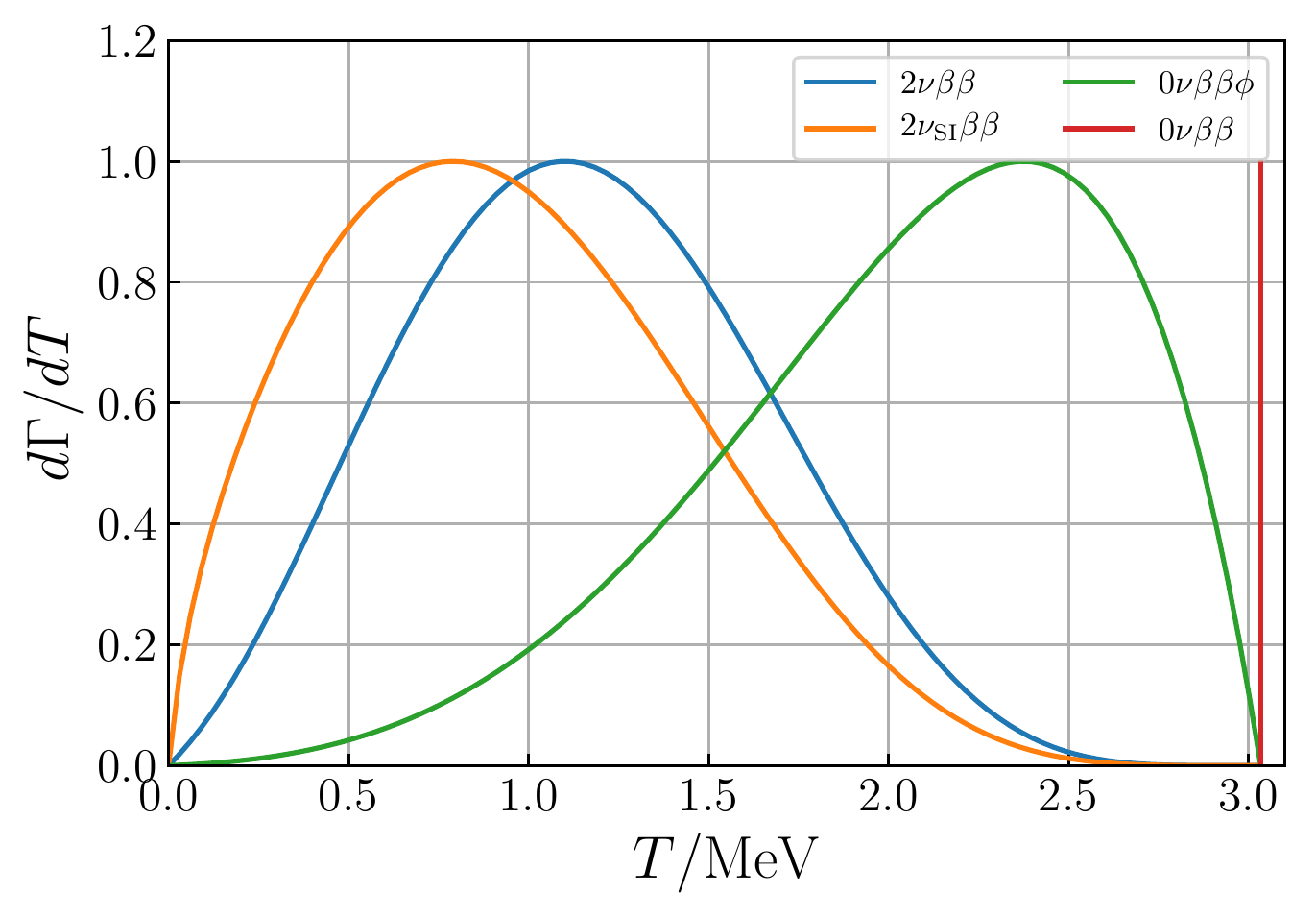}
	\caption{\label{fig:distortion}Spectra of \vSIbb{}, $2\nu\beta\beta$, $0\nu\beta\beta$ and $0\nu\beta\beta\phi$ decay with respect to the total electron kinetic energy $T$ for $^{100}$Mo. The \vSIbb{} spectrum is calculated for an $s$-channel mediator mass $m_\phi = Q + 0.1m_e$ and all spectra have been normalized to the same maximal height. 
	}
\end{figure}
In Fig.~\ref{fig:distortion}, we show the electron energy distributions of \vSIbb{} and $2\nu\beta\beta$ decay for the isotope $^{100}$Mo with an $s$-channel mediator mass $m_\phi = Q + 0.1m_e$, slightly above the kinematic threshold. For comparison, we also show a vertical line corresponding to $0\nu\beta\beta$ decay, and the distribution for Majoron emission ($0\nu\beta\beta\phi$) taken from Ref.~\cite{Brune:2018sab}. As can be seen in Fig.~\ref{fig:distortion}, the energy spectrum of \vSIbb{} decay is shifted towards lower energies when compared to the $2\nu\beta\beta$ spectrum. The shift can be understood qualitatively. With increasing $T$ the energy taken away by neutrinos is smaller, leading to a smaller value of $s$ and hence a smaller value of the $s$-channel $G_S$. To determine the experimental sensitivity to such distortion, we have performed a simple $\chi^2$-fit to the NEMO-3 data~\cite{NEMO-3:2019gwo} as detailed in \suppmat~\ref{sec:nemo3-fitting}. We find that for $\Gamma_{\nu{\rm SI}} = r^2 \Gamma_{2\nu}$ with $r = 16\%$, the $\chi^2$-value is changed by $\Delta \chi^2 = 9$ ($3\sigma$), which implies that if the spectral distortion is taken into account, the bound on $G_S$ can be approximately improved by one order of magnitude. We emphasize that this applies for the specific mediator mass $m_\phi = Q + 0.1m_e$ and the sensitivity will decrease for larger masses.

%
\begin{figure}[t]
	\centering
	\includegraphics[width=0.95\columnwidth]{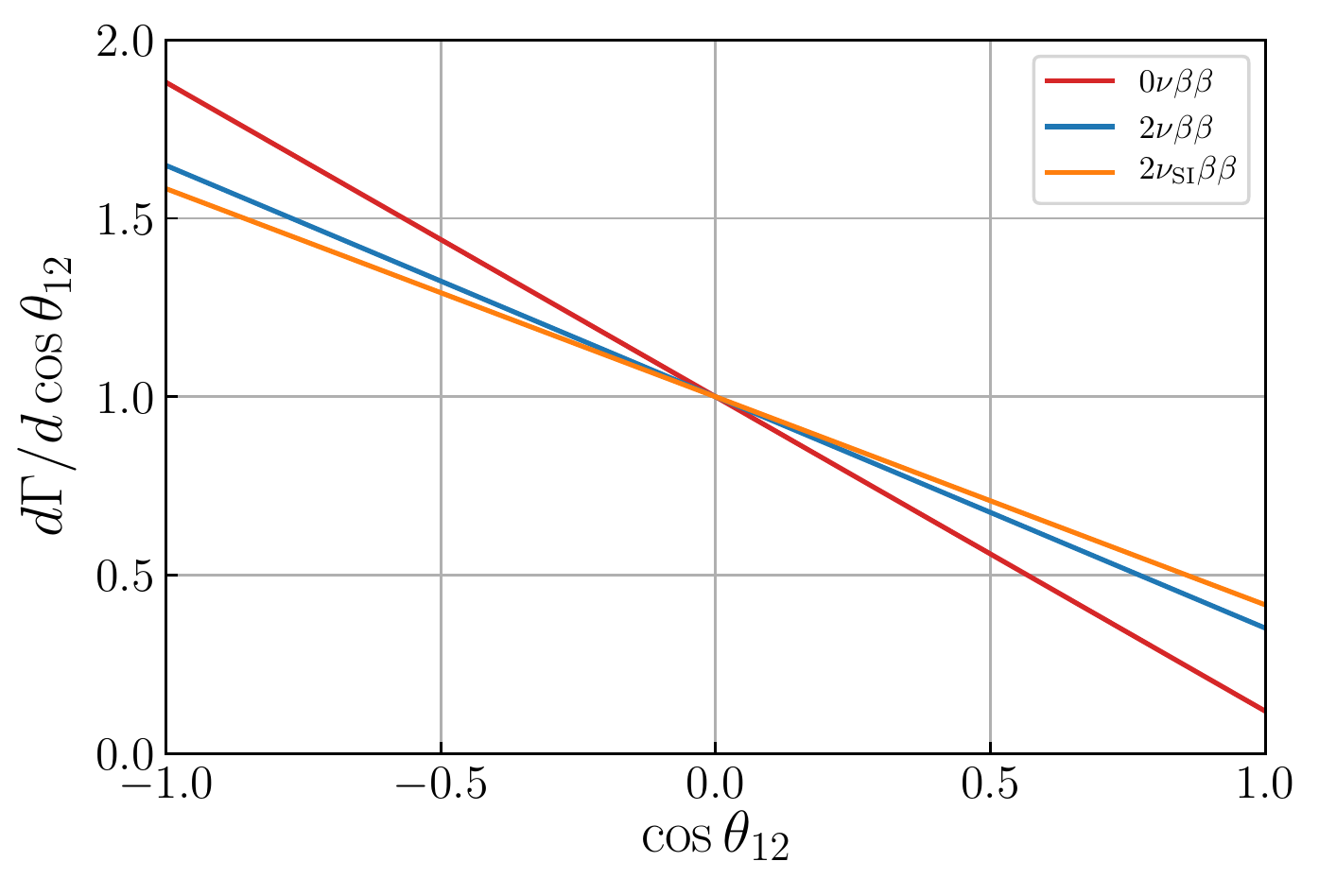}
	\caption{\label{fig:distortion-angular}As Fig.~\ref{fig:distortion}, but showing the electron angular distributions of \vSIbb{}, $2\nu\beta\beta$ and $0\nu\beta\beta$ decay. The lines have been normalized to a value $1$ at $\cos\theta_{12} = 0$, so the figure is in arbitrary units.}
\end{figure}

In addition to the energy distribution, the angular distribution can also be measured in dedicated experiments such as NEMO-3~\cite{NEMO-3:2019gwo} and SuperNEMO~\cite{Arnold:2010tu}. From Eq.~\eqref{eq:s-5-1}, the angular distribution of \vSIbb{} decay is obtained by integrating out $p_1$ and $p_2$. The result takes the form
\begin{equation}
	\frac{d\Gamma_{\nu{\rm SI}}}{d\cos\theta_{12}} = 
	\frac{\Gamma_{\nu\text{SI}}}{2}\left(1 - k_\theta^{\nu{\rm SI}}\cos\theta_{12}\right),
\label{eq:s-14}
\end{equation}
where the angular correlation $k_\theta^{\nu{\rm SI}}$ is computed in  \suppmat{}~\ref{sec:appA-leptonic}. 
%
%
For $2\nu\beta\beta$ and $0\nu\beta\beta$ decay, the electron angular distributions take the same form as Eq.~\eqref{eq:s-14}, but with different angular coefficients $k_\theta$ \cite{Doi:1985dx, Arnold:2010tu}. We refer to those as $k_\theta^{0\nu}$ and $k_\theta^{2\nu}$, respectively. Their expressions are given in \suppmat{}~\ref{sec:appA-leptonic} as well. For $^{100}$Mo, the numerical values are $k_\theta^{\nu{\rm SI}} = 0.58$ (using again $m_\phi = Q + 0.1m_e$), $k_\theta^{2\nu} = 0.65$ and $k_\theta^{0\nu} = 0.88$. With these values, we show in Fig.~\ref{fig:distortion-angular} the angular distributions of electrons for the three processes. Again, we have performed a $\chi^2$-fit to the NEMO-3 data~\cite{NEMO-3:2019gwo} and the result is $r < 29\%$ at 3$\sigma$ confidence level. This indicates that the angular distribution is less sensitive than the energy distribution to distortions from \vSIbb. 
This is in fact interesting, as among the running and future experiments only one (SuperNEMO) has sensitivity on the angular distribution.

\section{Conclusion and Discussion}
\label{sec:concl}
\noindent 
The search for $0\nu\beta\beta$ decay constitutes one of the most important aspects to determine the nature and properties of neutrinos. 
As we have demonstrated in Figs.~\ref{fig:limit}, \ref{fig:distortion}, and \ref{fig:distortion-angular}, in the presence of $\nu$SI involving two $\nu_e$, there can be significant effects not only on the total rates of $2\nu\beta\beta$ decay, but also on the spectrum shapes. If only the total rates are considered, we find that $^{136}$Xe currently has the best sensitivity to $\nu$SI. The observed $2\nu\beta\beta$ rate implies $G_S < (0.32 - 0.43) \times 10^9\, G_F$, which is significantly lower than the cosmologically favoured value $G_S = 3.83_{-0.54}^{+1.22}\times 10^9\, G_F$ in the strongly interacting regime. However, one should note that this bound does not apply if only $\nu_\mu$ and $\nu_\tau$ participate in $\nu$SI.

Including spectral distortions could further improve the sensitivities. This is of interest when 
the particle that mediates the self-interactions has a mass that is larger than the available $Q$-value of the double beta decay. 
The distortions are caused by the energy dependence of the effective coupling $G_S$ and hence are affected by the underlying models for $\nu$SI. In this work, we only consider an $s$-channel mediating scalar, which allows us to evade nuclear physics calculations and to quantitatively show spectral distortions of the energy and angular distributions. For other possibilities containing a $t$-channel dependence, very different spectral distortions could appear, which will be addressed when a more dedicated study involving nuclear physics calculations is carried out.   

In our calculations, we neglected the interference between the SM $2\nu\beta\beta$ and exotic \vSIbb{} contributions, which in principle would be present if two electron anti-neutrinos are emitted in the $\nu$SI-mediated process. We estimate that the interference contributes in \suppmat{}~\ref{sec:interference}. 
When 
%
the theoretical determination of the $2\nu\beta\beta$ rate becomes more precise, it will be important to include this interference term, but currently it does not improve the sensitivity to $G_S$. 

In summary, our work shows that strong $\nu$SI favored by the cosmological data might have an impact on  $2\nu\beta\beta$ decay experiments. Precision measurements of $2\nu\beta\beta$ decay spectra combined with more theoretical effort in computing NMEs have the potential of probing hidden interactions of neutrinos. Furthermore, it demonstrates the importance of having access to energy and angular distributions of electrons in double beta decay experiments.

\begin{acknowledgments}
\noindent 
W.R.\ is supported by the DFG with grant RO 2516/7-1 in the Heisenberg program. F.F.D.\ acknowledges support from the UK Science and Technology Facilities Council (STFC) via a Consolidated Grant (Reference ST/P00072X/1).
\end{acknowledgments}

\bibliographystyle{apsrev4-1}
\bibliography{ref}

\begin{thebibliography}{60}%
\makeatletter
\providecommand \@ifxundefined [1]{%
 \@ifx{#1\undefined}
}%
\providecommand \@ifnum [1]{%
 \ifnum #1\expandafter \@firstoftwo
 \else \expandafter \@secondoftwo
 \fi
}%
\providecommand \@ifx [1]{%
 \ifx #1\expandafter \@firstoftwo
 \else \expandafter \@secondoftwo
 \fi
}%
\providecommand \natexlab [1]{#1}%
\providecommand \enquote  [1]{``#1''}%
\providecommand \bibnamefont  [1]{#1}%
\providecommand \bibfnamefont [1]{#1}%
\providecommand \citenamefont [1]{#1}%
\providecommand \href@noop [0]{\@secondoftwo}%
\providecommand \href [0]{\begingroup \@sanitize@url \@href}%
\providecommand \@href[1]{\@@startlink{#1}\@@href}%
\providecommand \@@href[1]{\endgroup#1\@@endlink}%
\providecommand \@sanitize@url [0]{\catcode `\\12\catcode `\$12\catcode
  `\&12\catcode `\#12\catcode `\^12\catcode `\_12\catcode `\%12\relax}%
\providecommand \@@startlink[1]{}%
\providecommand \@@endlink[0]{}%
\providecommand \url  [0]{\begingroup\@sanitize@url \@url }%
\providecommand \@url [1]{\endgroup\@href {#1}{\urlprefix }}%
\providecommand \urlprefix  [0]{URL }%
\providecommand \Eprint [0]{\href }%
\providecommand \doibase [0]{http://dx.doi.org/}%
\providecommand \selectlanguage [0]{\@gobble}%
\providecommand \bibinfo  [0]{\@secondoftwo}%
\providecommand \bibfield  [0]{\@secondoftwo}%
\providecommand \translation [1]{[#1]}%
\providecommand \BibitemOpen [0]{}%
\providecommand \bibitemStop [0]{}%
\providecommand \bibitemNoStop [0]{.\EOS\space}%
\providecommand \EOS [0]{\spacefactor3000\relax}%
\providecommand \BibitemShut  [1]{\csname bibitem#1\endcsname}%
\let\auto@bib@innerbib\@empty
\bibitem [{\citenamefont {Riess}\ \emph {et~al.}(2016)\citenamefont {Riess}
  \emph {et~al.}}]{Riess:2016jrr}%
  \BibitemOpen
  \bibfield  {author} {\bibinfo {author} {\bibfnamefont {A.~G.}\ \bibnamefont
  {Riess}} \emph {et~al.},\ }\href {\doibase 10.3847/0004-637X/826/1/56}
  {\bibfield  {journal} {\bibinfo  {journal} {Astrophys. J.}\ }\textbf
  {\bibinfo {volume} {826}},\ \bibinfo {pages} {56} (\bibinfo {year} {2016})},\
  \Eprint {http://arxiv.org/abs/1604.01424} {arXiv:1604.01424 [astro-ph.CO]}
  \BibitemShut {NoStop}%
\bibitem [{\citenamefont {Shanks}\ \emph {et~al.}(2019)\citenamefont {Shanks},
  \citenamefont {Hogarth},\ and\ \citenamefont {Metcalfe}}]{Shanks:2018rka}%
  \BibitemOpen
  \bibfield  {author} {\bibinfo {author} {\bibfnamefont {T.}~\bibnamefont
  {Shanks}}, \bibinfo {author} {\bibfnamefont {L.}~\bibnamefont {Hogarth}}, \
  and\ \bibinfo {author} {\bibfnamefont {N.}~\bibnamefont {Metcalfe}},\ }\href
  {\doibase 10.1093/mnrasl/sly239} {\bibfield  {journal} {\bibinfo  {journal}
  {Mon. Not. Roy. Astron. Soc.}\ }\textbf {\bibinfo {volume} {484}},\ \bibinfo
  {pages} {L64} (\bibinfo {year} {2019})},\ \Eprint
  {http://arxiv.org/abs/1810.02595} {arXiv:1810.02595 [astro-ph.CO]}
  \BibitemShut {NoStop}%
\bibitem [{\citenamefont {Riess}\ \emph {et~al.}(2018)\citenamefont {Riess},
  \citenamefont {Casertano}, \citenamefont {Kenworthy}, \citenamefont
  {Scolnic},\ and\ \citenamefont {Macri}}]{Riess:2018kzi}%
  \BibitemOpen
  \bibfield  {author} {\bibinfo {author} {\bibfnamefont {A.~G.}\ \bibnamefont
  {Riess}}, \bibinfo {author} {\bibfnamefont {S.}~\bibnamefont {Casertano}},
  \bibinfo {author} {\bibfnamefont {D.}~\bibnamefont {Kenworthy}}, \bibinfo
  {author} {\bibfnamefont {D.}~\bibnamefont {Scolnic}}, \ and\ \bibinfo
  {author} {\bibfnamefont {L.}~\bibnamefont {Macri}},\ }\href@noop {} {\
  (\bibinfo {year} {2018})},\ \Eprint {http://arxiv.org/abs/1810.03526}
  {arXiv:1810.03526 [astro-ph.CO]} \BibitemShut {NoStop}%
\bibitem [{\citenamefont {Aghanim}\ \emph {et~al.}(2018)\citenamefont {Aghanim}
  \emph {et~al.}}]{Aghanim:2018eyx}%
  \BibitemOpen
  \bibfield  {author} {\bibinfo {author} {\bibfnamefont {N.}~\bibnamefont
  {Aghanim}} \emph {et~al.} (\bibinfo {collaboration} {Planck}),\ }\href@noop
  {} {\  (\bibinfo {year} {2018})},\ \Eprint {http://arxiv.org/abs/1807.06209}
  {arXiv:1807.06209 [astro-ph.CO]} \BibitemShut {NoStop}%
\bibitem [{\citenamefont {Riess}\ \emph {et~al.}(2019)\citenamefont {Riess},
  \citenamefont {Casertano}, \citenamefont {Yuan}, \citenamefont {Macri},\ and\
  \citenamefont {Scolnic}}]{Riess:2019cxk}%
  \BibitemOpen
  \bibfield  {author} {\bibinfo {author} {\bibfnamefont {A.~G.}\ \bibnamefont
  {Riess}}, \bibinfo {author} {\bibfnamefont {S.}~\bibnamefont {Casertano}},
  \bibinfo {author} {\bibfnamefont {W.}~\bibnamefont {Yuan}}, \bibinfo {author}
  {\bibfnamefont {L.~M.}\ \bibnamefont {Macri}}, \ and\ \bibinfo {author}
  {\bibfnamefont {D.}~\bibnamefont {Scolnic}},\ }\href {\doibase
  10.3847/1538-4357/ab1422} {\bibfield  {journal} {\bibinfo  {journal}
  {Astrophys. J.}\ }\textbf {\bibinfo {volume} {876}},\ \bibinfo {pages} {85}
  (\bibinfo {year} {2019})},\ \Eprint {http://arxiv.org/abs/1903.07603}
  {arXiv:1903.07603 [astro-ph.CO]} \BibitemShut {NoStop}%
\bibitem [{\citenamefont {Cyr-Racine}\ and\ \citenamefont
  {Sigurdson}(2014)}]{Cyr-Racine:2013jua}%
  \BibitemOpen
  \bibfield  {author} {\bibinfo {author} {\bibfnamefont {F.-Y.}\ \bibnamefont
  {Cyr-Racine}}\ and\ \bibinfo {author} {\bibfnamefont {K.}~\bibnamefont
  {Sigurdson}},\ }\href {\doibase 10.1103/PhysRevD.90.123533} {\bibfield
  {journal} {\bibinfo  {journal} {Phys. Rev.}\ }\textbf {\bibinfo {volume}
  {D90}},\ \bibinfo {pages} {123533} (\bibinfo {year} {2014})},\ \Eprint
  {http://arxiv.org/abs/1306.1536} {arXiv:1306.1536 [astro-ph.CO]} \BibitemShut
  {NoStop}%
\bibitem [{\citenamefont {Lancaster}\ \emph {et~al.}(2017)\citenamefont
  {Lancaster}, \citenamefont {Cyr-Racine}, \citenamefont {Knox},\ and\
  \citenamefont {Pan}}]{Lancaster:2017ksf}%
  \BibitemOpen
  \bibfield  {author} {\bibinfo {author} {\bibfnamefont {L.}~\bibnamefont
  {Lancaster}}, \bibinfo {author} {\bibfnamefont {F.-Y.}\ \bibnamefont
  {Cyr-Racine}}, \bibinfo {author} {\bibfnamefont {L.}~\bibnamefont {Knox}}, \
  and\ \bibinfo {author} {\bibfnamefont {Z.}~\bibnamefont {Pan}},\ }\href
  {\doibase 10.1088/1475-7516/2017/07/033} {\bibfield  {journal} {\bibinfo
  {journal} {JCAP}\ }\textbf {\bibinfo {volume} {1707}},\ \bibinfo {pages}
  {033} (\bibinfo {year} {2017})},\ \Eprint {http://arxiv.org/abs/1704.06657}
  {arXiv:1704.06657 [astro-ph.CO]} \BibitemShut {NoStop}%
\bibitem [{\citenamefont {Oldengott}\ \emph {et~al.}(2017)\citenamefont
  {Oldengott}, \citenamefont {Tram}, \citenamefont {Rampf},\ and\ \citenamefont
  {Wong}}]{Oldengott:2017fhy}%
  \BibitemOpen
  \bibfield  {author} {\bibinfo {author} {\bibfnamefont {I.~M.}\ \bibnamefont
  {Oldengott}}, \bibinfo {author} {\bibfnamefont {T.}~\bibnamefont {Tram}},
  \bibinfo {author} {\bibfnamefont {C.}~\bibnamefont {Rampf}}, \ and\ \bibinfo
  {author} {\bibfnamefont {Y.~Y.~Y.}\ \bibnamefont {Wong}},\ }\href {\doibase
  10.1088/1475-7516/2017/11/027} {\bibfield  {journal} {\bibinfo  {journal}
  {JCAP}\ }\textbf {\bibinfo {volume} {1711}},\ \bibinfo {pages} {027}
  (\bibinfo {year} {2017})},\ \Eprint {http://arxiv.org/abs/1706.02123}
  {arXiv:1706.02123 [astro-ph.CO]} \BibitemShut {NoStop}%
\bibitem [{\citenamefont {Kreisch}\ \emph {et~al.}(2019)\citenamefont
  {Kreisch}, \citenamefont {Cyr-Racine},\ and\ \citenamefont
  {Dore}}]{Kreisch:2019yzn}%
  \BibitemOpen
  \bibfield  {author} {\bibinfo {author} {\bibfnamefont {C.~D.}\ \bibnamefont
  {Kreisch}}, \bibinfo {author} {\bibfnamefont {F.-Y.}\ \bibnamefont
  {Cyr-Racine}}, \ and\ \bibinfo {author} {\bibfnamefont {O.}~\bibnamefont
  {Dore}},\ }\href@noop {} {\  (\bibinfo {year} {2019})},\ \Eprint
  {http://arxiv.org/abs/1902.00534} {arXiv:1902.00534 [astro-ph.CO]}
  \BibitemShut {NoStop}%
\bibitem [{\citenamefont {Park}\ \emph {et~al.}(2019)\citenamefont {Park},
  \citenamefont {Kreisch}, \citenamefont {Dunkley}, \citenamefont
  {Hadzhiyska},\ and\ \citenamefont {Cyr-Racine}}]{Park:2019ibn}%
  \BibitemOpen
  \bibfield  {author} {\bibinfo {author} {\bibfnamefont {M.}~\bibnamefont
  {Park}}, \bibinfo {author} {\bibfnamefont {C.~D.}\ \bibnamefont {Kreisch}},
  \bibinfo {author} {\bibfnamefont {J.}~\bibnamefont {Dunkley}}, \bibinfo
  {author} {\bibfnamefont {B.}~\bibnamefont {Hadzhiyska}}, \ and\ \bibinfo
  {author} {\bibfnamefont {F.-Y.}\ \bibnamefont {Cyr-Racine}},\ }\href
  {\doibase 10.1103/PhysRevD.100.063524} {\bibfield  {journal} {\bibinfo
  {journal} {Phys. Rev.}\ }\textbf {\bibinfo {volume} {D100}},\ \bibinfo
  {pages} {063524} (\bibinfo {year} {2019})},\ \Eprint
  {http://arxiv.org/abs/1904.02625} {arXiv:1904.02625 [astro-ph.CO]}
  \BibitemShut {NoStop}%
\bibitem [{\citenamefont {Hasenkamp}(2016)}]{Hasenkamp:2016pme}%
  \BibitemOpen
  \bibfield  {author} {\bibinfo {author} {\bibfnamefont {J.}~\bibnamefont
  {Hasenkamp}},\ }\href {\doibase 10.1103/PhysRevD.93.055033} {\bibfield
  {journal} {\bibinfo  {journal} {Phys. Rev.}\ }\textbf {\bibinfo {volume}
  {D93}},\ \bibinfo {pages} {055033} (\bibinfo {year} {2016})},\ \Eprint
  {http://arxiv.org/abs/1604.04742} {arXiv:1604.04742 [hep-ph]} \BibitemShut
  {NoStop}%
\bibitem [{\citenamefont {Huang}\ \emph {et~al.}(2018)\citenamefont {Huang},
  \citenamefont {Ohlsson},\ and\ \citenamefont {Zhou}}]{Huang:2017egl}%
  \BibitemOpen
  \bibfield  {author} {\bibinfo {author} {\bibfnamefont {G.-y.}\ \bibnamefont
  {Huang}}, \bibinfo {author} {\bibfnamefont {T.}~\bibnamefont {Ohlsson}}, \
  and\ \bibinfo {author} {\bibfnamefont {S.}~\bibnamefont {Zhou}},\ }\href
  {\doibase 10.1103/PhysRevD.97.075009} {\bibfield  {journal} {\bibinfo
  {journal} {Phys. Rev.}\ }\textbf {\bibinfo {volume} {D97}},\ \bibinfo {pages}
  {075009} (\bibinfo {year} {2018})},\ \Eprint
  {http://arxiv.org/abs/1712.04792} {arXiv:1712.04792 [hep-ph]} \BibitemShut
  {NoStop}%
\bibitem [{\citenamefont {Bakhti}\ \emph {et~al.}(2019)\citenamefont {Bakhti},
  \citenamefont {Farzan},\ and\ \citenamefont {Rajaee}}]{Bakhti:2018avv}%
  \BibitemOpen
  \bibfield  {author} {\bibinfo {author} {\bibfnamefont {P.}~\bibnamefont
  {Bakhti}}, \bibinfo {author} {\bibfnamefont {Y.}~\bibnamefont {Farzan}}, \
  and\ \bibinfo {author} {\bibfnamefont {M.}~\bibnamefont {Rajaee}},\ }\href
  {\doibase 10.1103/PhysRevD.99.055019} {\bibfield  {journal} {\bibinfo
  {journal} {Phys. Rev.}\ }\textbf {\bibinfo {volume} {D99}},\ \bibinfo {pages}
  {055019} (\bibinfo {year} {2019})},\ \Eprint
  {http://arxiv.org/abs/1810.04441} {arXiv:1810.04441 [hep-ph]} \BibitemShut
  {NoStop}%
\bibitem [{\citenamefont {Blinov}\ \emph {et~al.}(2019)\citenamefont {Blinov},
  \citenamefont {Kelly}, \citenamefont {Krnjaic},\ and\ \citenamefont
  {McDermott}}]{Blinov:2019gcj}%
  \BibitemOpen
  \bibfield  {author} {\bibinfo {author} {\bibfnamefont {N.}~\bibnamefont
  {Blinov}}, \bibinfo {author} {\bibfnamefont {K.~J.}\ \bibnamefont {Kelly}},
  \bibinfo {author} {\bibfnamefont {G.~Z.}\ \bibnamefont {Krnjaic}}, \ and\
  \bibinfo {author} {\bibfnamefont {S.~D.}\ \bibnamefont {McDermott}},\ }\href
  {\doibase 10.1103/PhysRevLett.123.191102} {\bibfield  {journal} {\bibinfo
  {journal} {Phys. Rev. Lett.}\ }\textbf {\bibinfo {volume} {123}},\ \bibinfo
  {pages} {191102} (\bibinfo {year} {2019})},\ \Eprint
  {http://arxiv.org/abs/1905.02727} {arXiv:1905.02727 [astro-ph.CO]}
  \BibitemShut {NoStop}%
\bibitem [{\citenamefont {De~Gouvea}\ \emph {et~al.}(2019)\citenamefont
  {De~Gouvea}, \citenamefont {Sen}, \citenamefont {Tangarife},\ and\
  \citenamefont {Zhang}}]{deGouvea:2019phk}%
  \BibitemOpen
  \bibfield  {author} {\bibinfo {author} {\bibfnamefont {A.}~\bibnamefont
  {De~Gouvea}}, \bibinfo {author} {\bibfnamefont {M.}~\bibnamefont {Sen}},
  \bibinfo {author} {\bibfnamefont {W.}~\bibnamefont {Tangarife}}, \ and\
  \bibinfo {author} {\bibfnamefont {Y.}~\bibnamefont {Zhang}},\ }\href@noop {}
  {\  (\bibinfo {year} {2019})},\ \Eprint {http://arxiv.org/abs/1910.04901}
  {arXiv:1910.04901 [hep-ph]} \BibitemShut {NoStop}%
\bibitem [{\citenamefont {Das}\ \emph {et~al.}(2017)\citenamefont {Das},
  \citenamefont {Dighe},\ and\ \citenamefont {Sen}}]{Das:2017iuj}%
  \BibitemOpen
  \bibfield  {author} {\bibinfo {author} {\bibfnamefont {A.}~\bibnamefont
  {Das}}, \bibinfo {author} {\bibfnamefont {A.}~\bibnamefont {Dighe}}, \ and\
  \bibinfo {author} {\bibfnamefont {M.}~\bibnamefont {Sen}},\ }\href {\doibase
  10.1088/1475-7516/2017/05/051} {\bibfield  {journal} {\bibinfo  {journal}
  {JCAP}\ }\textbf {\bibinfo {volume} {1705}},\ \bibinfo {pages} {051}
  (\bibinfo {year} {2017})},\ \Eprint {http://arxiv.org/abs/1705.00468}
  {arXiv:1705.00468 [hep-ph]} \BibitemShut {NoStop}%
\bibitem [{\citenamefont {Dighe}\ and\ \citenamefont
  {Sen}(2018)}]{Dighe:2017sur}%
  \BibitemOpen
  \bibfield  {author} {\bibinfo {author} {\bibfnamefont {A.}~\bibnamefont
  {Dighe}}\ and\ \bibinfo {author} {\bibfnamefont {M.}~\bibnamefont {Sen}},\
  }\href {\doibase 10.1103/PhysRevD.97.043011} {\bibfield  {journal} {\bibinfo
  {journal} {Phys. Rev.}\ }\textbf {\bibinfo {volume} {D97}},\ \bibinfo {pages}
  {043011} (\bibinfo {year} {2018})},\ \Eprint
  {http://arxiv.org/abs/1709.06858} {arXiv:1709.06858 [hep-ph]} \BibitemShut
  {NoStop}%
\bibitem [{\citenamefont {Ko}\ \emph {et~al.}(2020)\citenamefont {Ko} \emph
  {et~al.}}]{Ko:2019ssx}%
  \BibitemOpen
  \bibfield  {author} {\bibinfo {author} {\bibfnamefont {H.}~\bibnamefont {Ko}}
  \emph {et~al.},\ }\bibfield  {booktitle} {\emph {\bibinfo {booktitle}
  {{Proceedings, 15th International Symposium on Origin of Matter and Evolution
  of the Galaxies (OMEG15): Kyoto, Japan}}},\ }\href {\doibase
  10.7566/JPSCP.31.011027} {\bibfield  {journal} {\bibinfo  {journal} {JPS
  Conf. Proc.}\ }\textbf {\bibinfo {volume} {31}},\ \bibinfo {pages} {011027}
  (\bibinfo {year} {2020})},\ \Eprint {http://arxiv.org/abs/1903.02086}
  {arXiv:1903.02086 [astro-ph.HE]} \BibitemShut {NoStop}%
\bibitem [{\citenamefont {Shalgar}\ \emph {et~al.}(2019)\citenamefont
  {Shalgar}, \citenamefont {Tamborra},\ and\ \citenamefont
  {Bustamante}}]{Shalgar:2019rqe}%
  \BibitemOpen
  \bibfield  {author} {\bibinfo {author} {\bibfnamefont {S.}~\bibnamefont
  {Shalgar}}, \bibinfo {author} {\bibfnamefont {I.}~\bibnamefont {Tamborra}}, \
  and\ \bibinfo {author} {\bibfnamefont {M.}~\bibnamefont {Bustamante}},\
  }\href@noop {} {\  (\bibinfo {year} {2019})},\ \Eprint
  {http://arxiv.org/abs/1912.09115} {arXiv:1912.09115 [astro-ph.HE]}
  \BibitemShut {NoStop}%
\bibitem [{\citenamefont {Forastieri}\ \emph {et~al.}(2019)\citenamefont
  {Forastieri}, \citenamefont {Lattanzi},\ and\ \citenamefont
  {Natoli}}]{Forastieri:2019cuf}%
  \BibitemOpen
  \bibfield  {author} {\bibinfo {author} {\bibfnamefont {F.}~\bibnamefont
  {Forastieri}}, \bibinfo {author} {\bibfnamefont {M.}~\bibnamefont
  {Lattanzi}}, \ and\ \bibinfo {author} {\bibfnamefont {P.}~\bibnamefont
  {Natoli}},\ }\href {\doibase 10.1103/PhysRevD.100.103526} {\bibfield
  {journal} {\bibinfo  {journal} {Phys. Rev.}\ }\textbf {\bibinfo {volume}
  {D100}},\ \bibinfo {pages} {103526} (\bibinfo {year} {2019})},\ \Eprint
  {http://arxiv.org/abs/1904.07810} {arXiv:1904.07810 [astro-ph.CO]}
  \BibitemShut {NoStop}%
\bibitem [{\citenamefont {Lyu}\ \emph {et~al.}(2020)\citenamefont {Lyu},
  \citenamefont {Stamou},\ and\ \citenamefont {Wang}}]{Lyu:2020lps}%
  \BibitemOpen
  \bibfield  {author} {\bibinfo {author} {\bibfnamefont {K.-F.}\ \bibnamefont
  {Lyu}}, \bibinfo {author} {\bibfnamefont {E.}~\bibnamefont {Stamou}}, \ and\
  \bibinfo {author} {\bibfnamefont {L.-T.}\ \bibnamefont {Wang}},\ }\href@noop
  {} {\  (\bibinfo {year} {2020})},\ \Eprint {http://arxiv.org/abs/2004.10868}
  {arXiv:2004.10868 [hep-ph]} \BibitemShut {NoStop}%
\bibitem [{\citenamefont {Boehm}\ \emph {et~al.}(2012)\citenamefont {Boehm},
  \citenamefont {Dolan},\ and\ \citenamefont {McCabe}}]{Boehm:2012gr}%
  \BibitemOpen
  \bibfield  {author} {\bibinfo {author} {\bibfnamefont {C.}~\bibnamefont
  {Boehm}}, \bibinfo {author} {\bibfnamefont {M.~J.}\ \bibnamefont {Dolan}}, \
  and\ \bibinfo {author} {\bibfnamefont {C.}~\bibnamefont {McCabe}},\ }\href
  {\doibase 10.1088/1475-7516/2012/12/027} {\bibfield  {journal} {\bibinfo
  {journal} {JCAP}\ }\textbf {\bibinfo {volume} {1212}},\ \bibinfo {pages}
  {027} (\bibinfo {year} {2012})},\ \Eprint {http://arxiv.org/abs/1207.0497}
  {arXiv:1207.0497 [astro-ph.CO]} \BibitemShut {NoStop}%
\bibitem [{\citenamefont {Kamada}\ and\ \citenamefont
  {Yu}(2015)}]{Kamada:2015era}%
  \BibitemOpen
  \bibfield  {author} {\bibinfo {author} {\bibfnamefont {A.}~\bibnamefont
  {Kamada}}\ and\ \bibinfo {author} {\bibfnamefont {H.-B.}\ \bibnamefont
  {Yu}},\ }\href {\doibase 10.1103/PhysRevD.92.113004} {\bibfield  {journal}
  {\bibinfo  {journal} {Phys. Rev.}\ }\textbf {\bibinfo {volume} {D92}},\
  \bibinfo {pages} {113004} (\bibinfo {year} {2015})},\ \Eprint
  {http://arxiv.org/abs/1504.00711} {arXiv:1504.00711 [hep-ph]} \BibitemShut
  {NoStop}%
\bibitem [{\citenamefont {Barger}\ \emph {et~al.}(1982)\citenamefont {Barger},
  \citenamefont {Keung},\ and\ \citenamefont {Pakvasa}}]{Barger:1981vd}%
  \BibitemOpen
  \bibfield  {author} {\bibinfo {author} {\bibfnamefont {V.~D.}\ \bibnamefont
  {Barger}}, \bibinfo {author} {\bibfnamefont {W.-Y.}\ \bibnamefont {Keung}}, \
  and\ \bibinfo {author} {\bibfnamefont {S.}~\bibnamefont {Pakvasa}},\ }\href
  {\doibase 10.1103/PhysRevD.25.907} {\bibfield  {journal} {\bibinfo  {journal}
  {Phys. Rev.}\ }\textbf {\bibinfo {volume} {D25}},\ \bibinfo {pages} {907}
  (\bibinfo {year} {1982})}\BibitemShut {NoStop}%
\bibitem [{\citenamefont {Lessa}\ and\ \citenamefont
  {Peres}(2007)}]{Lessa:2007up}%
  \BibitemOpen
  \bibfield  {author} {\bibinfo {author} {\bibfnamefont {A.~P.}\ \bibnamefont
  {Lessa}}\ and\ \bibinfo {author} {\bibfnamefont {O.~L.~G.}\ \bibnamefont
  {Peres}},\ }\href {\doibase 10.1103/PhysRevD.75.094001} {\bibfield  {journal}
  {\bibinfo  {journal} {Phys. Rev.}\ }\textbf {\bibinfo {volume} {D75}},\
  \bibinfo {pages} {094001} (\bibinfo {year} {2007})},\ \Eprint
  {http://arxiv.org/abs/hep-ph/0701068} {arXiv:hep-ph/0701068 [hep-ph]}
  \BibitemShut {NoStop}%
\bibitem [{\citenamefont {Pasquini}\ and\ \citenamefont
  {Peres}(2016)}]{Pasquini:2015fjv}%
  \BibitemOpen
  \bibfield  {author} {\bibinfo {author} {\bibfnamefont {P.~S.}\ \bibnamefont
  {Pasquini}}\ and\ \bibinfo {author} {\bibfnamefont {O.~L.~G.}\ \bibnamefont
  {Peres}},\ }\href {\doibase 10.1103/PhysRevD.93.053007,
  10.1103/PhysRevD.93.079902} {\bibfield  {journal} {\bibinfo  {journal} {Phys.
  Rev.}\ }\textbf {\bibinfo {volume} {D93}},\ \bibinfo {pages} {053007}
  (\bibinfo {year} {2016})},\ \bibinfo {note} {[Erratum: Phys.
  Rev.D93,no.7,079902(2016)]},\ \Eprint {http://arxiv.org/abs/1511.01811}
  {arXiv:1511.01811 [hep-ph]} \BibitemShut {NoStop}%
\bibitem [{\citenamefont {Berryman}\ \emph {et~al.}(2018)\citenamefont
  {Berryman}, \citenamefont {De~Gouvea}, \citenamefont {Kelly},\ and\
  \citenamefont {Zhang}}]{Berryman:2018ogk}%
  \BibitemOpen
  \bibfield  {author} {\bibinfo {author} {\bibfnamefont {J.~M.}\ \bibnamefont
  {Berryman}}, \bibinfo {author} {\bibfnamefont {A.}~\bibnamefont {De~Gouvea}},
  \bibinfo {author} {\bibfnamefont {K.~J.}\ \bibnamefont {Kelly}}, \ and\
  \bibinfo {author} {\bibfnamefont {Y.}~\bibnamefont {Zhang}},\ }\href
  {\doibase 10.1103/PhysRevD.97.075030} {\bibfield  {journal} {\bibinfo
  {journal} {Phys. Rev.}\ }\textbf {\bibinfo {volume} {D97}},\ \bibinfo {pages}
  {075030} (\bibinfo {year} {2018})},\ \Eprint
  {http://arxiv.org/abs/1802.00009} {arXiv:1802.00009 [hep-ph]} \BibitemShut
  {NoStop}%
\bibitem [{\citenamefont {Brdar}\ \emph {et~al.}(2020)\citenamefont {Brdar},
  \citenamefont {Lindner}, \citenamefont {Vogl},\ and\ \citenamefont
  {Xu}}]{Brdar:2020nbj}%
  \BibitemOpen
  \bibfield  {author} {\bibinfo {author} {\bibfnamefont {V.}~\bibnamefont
  {Brdar}}, \bibinfo {author} {\bibfnamefont {M.}~\bibnamefont {Lindner}},
  \bibinfo {author} {\bibfnamefont {S.}~\bibnamefont {Vogl}}, \ and\ \bibinfo
  {author} {\bibfnamefont {X.-J.}\ \bibnamefont {Xu}},\ }\href@noop {} {\
  (\bibinfo {year} {2020})},\ \Eprint {http://arxiv.org/abs/2003.05339}
  {arXiv:2003.05339 [hep-ph]} \BibitemShut {NoStop}%
\bibitem [{\citenamefont {Arcadi}\ \emph {et~al.}(2019)\citenamefont {Arcadi},
  \citenamefont {Heeck}, \citenamefont {Heizmann}, \citenamefont {Mertens},
  \citenamefont {Queiroz}, \citenamefont {Rodejohann}, \citenamefont {Slezak},\
  and\ \citenamefont {Valerius}}]{Arcadi:2018xdd}%
  \BibitemOpen
  \bibfield  {author} {\bibinfo {author} {\bibfnamefont {G.}~\bibnamefont
  {Arcadi}}, \bibinfo {author} {\bibfnamefont {J.}~\bibnamefont {Heeck}},
  \bibinfo {author} {\bibfnamefont {F.}~\bibnamefont {Heizmann}}, \bibinfo
  {author} {\bibfnamefont {S.}~\bibnamefont {Mertens}}, \bibinfo {author}
  {\bibfnamefont {F.~S.}\ \bibnamefont {Queiroz}}, \bibinfo {author}
  {\bibfnamefont {W.}~\bibnamefont {Rodejohann}}, \bibinfo {author}
  {\bibfnamefont {M.}~\bibnamefont {Slezak}}, \ and\ \bibinfo {author}
  {\bibfnamefont {K.}~\bibnamefont {Valerius}},\ }\href {\doibase
  10.1007/JHEP01(2019)206} {\bibfield  {journal} {\bibinfo  {journal} {JHEP}\
  }\textbf {\bibinfo {volume} {01}},\ \bibinfo {pages} {206} (\bibinfo {year}
  {2019})},\ \Eprint {http://arxiv.org/abs/1811.03530} {arXiv:1811.03530
  [hep-ph]} \BibitemShut {NoStop}%
\bibitem [{\citenamefont {Deppisch}\ \emph {et~al.}(2012)\citenamefont
  {Deppisch}, \citenamefont {Hirsch},\ and\ \citenamefont
  {Pas}}]{Deppisch:2012nb}%
  \BibitemOpen
  \bibfield  {author} {\bibinfo {author} {\bibfnamefont {F.~F.}\ \bibnamefont
  {Deppisch}}, \bibinfo {author} {\bibfnamefont {M.}~\bibnamefont {Hirsch}}, \
  and\ \bibinfo {author} {\bibfnamefont {H.}~\bibnamefont {Pas}},\ }\href
  {\doibase 10.1088/0954-3899/39/12/124007} {\bibfield  {journal} {\bibinfo
  {journal} {J. Phys. G}\ }\textbf {\bibinfo {volume} {39}},\ \bibinfo {pages}
  {124007} (\bibinfo {year} {2012})},\ \Eprint {http://arxiv.org/abs/1208.0727}
  {arXiv:1208.0727 [hep-ph]} \BibitemShut {NoStop}%
\bibitem [{\citenamefont {Dolinski}\ \emph {et~al.}(2019)\citenamefont
  {Dolinski}, \citenamefont {Poon},\ and\ \citenamefont
  {Rodejohann}}]{Dolinski:2019nrj}%
  \BibitemOpen
  \bibfield  {author} {\bibinfo {author} {\bibfnamefont {M.~J.}\ \bibnamefont
  {Dolinski}}, \bibinfo {author} {\bibfnamefont {A.~W.~P.}\ \bibnamefont
  {Poon}}, \ and\ \bibinfo {author} {\bibfnamefont {W.}~\bibnamefont
  {Rodejohann}},\ }\href {\doibase 10.1146/annurev-nucl-101918-023407}
  {\bibfield  {journal} {\bibinfo  {journal} {Ann. Rev. Nucl. Part. Sci.}\
  }\textbf {\bibinfo {volume} {69}},\ \bibinfo {pages} {219} (\bibinfo {year}
  {2019})},\ \Eprint {http://arxiv.org/abs/1902.04097} {arXiv:1902.04097
  [nucl-ex]} \BibitemShut {NoStop}%
\bibitem [{\citenamefont {Deppisch}\ \emph {et~al.}(2020)\citenamefont
  {Deppisch}, \citenamefont {Graf},\ and\ \citenamefont
  {Simkovic}}]{Deppisch:2020mxv}%
  \BibitemOpen
  \bibfield  {author} {\bibinfo {author} {\bibfnamefont {F.~F.}\ \bibnamefont
  {Deppisch}}, \bibinfo {author} {\bibfnamefont {L.}~\bibnamefont {Graf}}, \
  and\ \bibinfo {author} {\bibfnamefont {F.}~\bibnamefont {Simkovic}},\
  }\href@noop {} {\  (\bibinfo {year} {2020})},\ \Eprint
  {http://arxiv.org/abs/2003.11836} {arXiv:2003.11836 [hep-ph]} \BibitemShut
  {NoStop}%
\bibitem [{\citenamefont {Burgess}\ and\ \citenamefont
  {Cline}(1993)}]{Burgess:1992dt}%
  \BibitemOpen
  \bibfield  {author} {\bibinfo {author} {\bibfnamefont {C.~P.}\ \bibnamefont
  {Burgess}}\ and\ \bibinfo {author} {\bibfnamefont {J.~M.}\ \bibnamefont
  {Cline}},\ }\href {\doibase 10.1016/0370-2693(93)91720-8} {\bibfield
  {journal} {\bibinfo  {journal} {Phys. Lett.}\ }\textbf {\bibinfo {volume}
  {B298}},\ \bibinfo {pages} {141} (\bibinfo {year} {1993})},\ \Eprint
  {http://arxiv.org/abs/hep-ph/9209299} {arXiv:hep-ph/9209299 [hep-ph]}
  \BibitemShut {NoStop}%
\bibitem [{\citenamefont {Burgess}\ and\ \citenamefont
  {Cline}(1994)}]{Burgess:1993xh}%
  \BibitemOpen
  \bibfield  {author} {\bibinfo {author} {\bibfnamefont {C.~P.}\ \bibnamefont
  {Burgess}}\ and\ \bibinfo {author} {\bibfnamefont {J.~M.}\ \bibnamefont
  {Cline}},\ }\href {\doibase 10.1103/PhysRevD.49.5925} {\bibfield  {journal}
  {\bibinfo  {journal} {Phys. Rev.}\ }\textbf {\bibinfo {volume} {D49}},\
  \bibinfo {pages} {5925} (\bibinfo {year} {1994})},\ \Eprint
  {http://arxiv.org/abs/hep-ph/9307316} {arXiv:hep-ph/9307316 [hep-ph]}
  \BibitemShut {NoStop}%
\bibitem [{\citenamefont {Gando}\ \emph {et~al.}(2012)\citenamefont {Gando}
  \emph {et~al.}}]{Gando:2012pj}%
  \BibitemOpen
  \bibfield  {author} {\bibinfo {author} {\bibfnamefont {A.}~\bibnamefont
  {Gando}} \emph {et~al.} (\bibinfo {collaboration} {KamLAND-Zen}),\ }\href
  {\doibase 10.1103/PhysRevC.86.021601} {\bibfield  {journal} {\bibinfo
  {journal} {Phys. Rev.}\ }\textbf {\bibinfo {volume} {C86}},\ \bibinfo {pages}
  {021601} (\bibinfo {year} {2012})},\ \Eprint {http://arxiv.org/abs/1205.6372}
  {arXiv:1205.6372 [hep-ex]} \BibitemShut {NoStop}%
\bibitem [{\citenamefont {Agostini}\ \emph {et~al.}(2015)\citenamefont
  {Agostini} \emph {et~al.}}]{Agostini:2015nwa}%
  \BibitemOpen
  \bibfield  {author} {\bibinfo {author} {\bibfnamefont {M.}~\bibnamefont
  {Agostini}} \emph {et~al.},\ }\href {\doibase 10.1140/epjc/s10052-015-3627-y}
  {\bibfield  {journal} {\bibinfo  {journal} {Eur. Phys. J.}\ }\textbf
  {\bibinfo {volume} {C75}},\ \bibinfo {pages} {416} (\bibinfo {year}
  {2015})},\ \Eprint {http://arxiv.org/abs/1501.02345} {arXiv:1501.02345
  [nucl-ex]} \BibitemShut {NoStop}%
\bibitem [{\citenamefont {Blum}\ \emph {et~al.}(2018)\citenamefont {Blum},
  \citenamefont {Nir},\ and\ \citenamefont {Shavit}}]{Blum:2018ljv}%
  \BibitemOpen
  \bibfield  {author} {\bibinfo {author} {\bibfnamefont {K.}~\bibnamefont
  {Blum}}, \bibinfo {author} {\bibfnamefont {Y.}~\bibnamefont {Nir}}, \ and\
  \bibinfo {author} {\bibfnamefont {M.}~\bibnamefont {Shavit}},\ }\href
  {\doibase 10.1016/j.physletb.2018.08.022} {\bibfield  {journal} {\bibinfo
  {journal} {Phys. Lett.}\ }\textbf {\bibinfo {volume} {B785}},\ \bibinfo
  {pages} {354} (\bibinfo {year} {2018})},\ \Eprint
  {http://arxiv.org/abs/1802.08019} {arXiv:1802.08019 [hep-ph]} \BibitemShut
  {NoStop}%
\bibitem [{\citenamefont {Cepedello}\ \emph {et~al.}(2019)\citenamefont
  {Cepedello}, \citenamefont {Deppisch}, \citenamefont {Gonzalez},
  \citenamefont {Hati},\ and\ \citenamefont {Hirsch}}]{Cepedello:2018zvr}%
  \BibitemOpen
  \bibfield  {author} {\bibinfo {author} {\bibfnamefont {R.}~\bibnamefont
  {Cepedello}}, \bibinfo {author} {\bibfnamefont {F.~F.}\ \bibnamefont
  {Deppisch}}, \bibinfo {author} {\bibfnamefont {L.}~\bibnamefont {Gonzalez}},
  \bibinfo {author} {\bibfnamefont {C.}~\bibnamefont {Hati}}, \ and\ \bibinfo
  {author} {\bibfnamefont {M.}~\bibnamefont {Hirsch}},\ }\href {\doibase
  10.1103/PhysRevLett.122.181801} {\bibfield  {journal} {\bibinfo  {journal}
  {Phys. Rev. Lett.}\ }\textbf {\bibinfo {volume} {122}},\ \bibinfo {pages}
  {181801} (\bibinfo {year} {2019})},\ \Eprint
  {http://arxiv.org/abs/1811.00031} {arXiv:1811.00031 [hep-ph]} \BibitemShut
  {NoStop}%
\bibitem [{\citenamefont {Brune}\ and\ \citenamefont
  {Paes}(2019)}]{Brune:2018sab}%
  \BibitemOpen
  \bibfield  {author} {\bibinfo {author} {\bibfnamefont {T.}~\bibnamefont
  {Brune}}\ and\ \bibinfo {author} {\bibfnamefont {H.}~\bibnamefont {Paes}},\
  }\href {\doibase 10.1103/PhysRevD.99.096005} {\bibfield  {journal} {\bibinfo
  {journal} {Phys. Rev.}\ }\textbf {\bibinfo {volume} {D99}},\ \bibinfo {pages}
  {096005} (\bibinfo {year} {2019})},\ \Eprint
  {http://arxiv.org/abs/1808.08158} {arXiv:1808.08158 [hep-ph]} \BibitemShut
  {NoStop}%
\bibitem [{\citenamefont {Farzan}\ \emph {et~al.}(2018)\citenamefont {Farzan},
  \citenamefont {Lindner}, \citenamefont {Rodejohann},\ and\ \citenamefont
  {Xu}}]{Farzan:2018gtr}%
  \BibitemOpen
  \bibfield  {author} {\bibinfo {author} {\bibfnamefont {Y.}~\bibnamefont
  {Farzan}}, \bibinfo {author} {\bibfnamefont {M.}~\bibnamefont {Lindner}},
  \bibinfo {author} {\bibfnamefont {W.}~\bibnamefont {Rodejohann}}, \ and\
  \bibinfo {author} {\bibfnamefont {X.-J.}\ \bibnamefont {Xu}},\ }\href
  {\doibase 10.1007/JHEP05(2018)066} {\bibfield  {journal} {\bibinfo  {journal}
  {JHEP}\ }\textbf {\bibinfo {volume} {05}},\ \bibinfo {pages} {066} (\bibinfo
  {year} {2018})},\ \Eprint {http://arxiv.org/abs/1802.05171} {arXiv:1802.05171
  [hep-ph]} \BibitemShut {NoStop}%
\bibitem [{\citenamefont {Barabash}(2019)}]{Barabash:2019nnr}%
  \BibitemOpen
  \bibfield  {author} {\bibinfo {author} {\bibfnamefont {A.}~\bibnamefont
  {Barabash}},\ }\href {\doibase 10.1063/1.5130963} {\bibfield  {journal}
  {\bibinfo  {journal} {AIP Conf. Proc.}\ }\textbf {\bibinfo {volume} {2165}},\
  \bibinfo {pages} {020002} (\bibinfo {year} {2019})},\ \Eprint
  {http://arxiv.org/abs/1907.06887} {arXiv:1907.06887 [nucl-ex]} \BibitemShut
  {NoStop}%
\bibitem [{\citenamefont {Barea}\ \emph {et~al.}(2015)\citenamefont {Barea},
  \citenamefont {Kotila},\ and\ \citenamefont {Iachello}}]{Barea:2015kwa}%
  \BibitemOpen
  \bibfield  {author} {\bibinfo {author} {\bibfnamefont {J.}~\bibnamefont
  {Barea}}, \bibinfo {author} {\bibfnamefont {J.}~\bibnamefont {Kotila}}, \
  and\ \bibinfo {author} {\bibfnamefont {F.}~\bibnamefont {Iachello}},\ }\href
  {\doibase 10.1103/PhysRevC.91.034304} {\bibfield  {journal} {\bibinfo
  {journal} {Phys. Rev.}\ }\textbf {\bibinfo {volume} {C91}},\ \bibinfo {pages}
  {034304} (\bibinfo {year} {2015})},\ \Eprint
  {http://arxiv.org/abs/1506.08530} {arXiv:1506.08530 [nucl-th]} \BibitemShut
  {NoStop}%
\bibitem [{\citenamefont {Redshaw}\ \emph {et~al.}(2012)\citenamefont
  {Redshaw}, \citenamefont {Bollen}, \citenamefont {Brodeur}, \citenamefont
  {Bustabad}, \citenamefont {Lincoln}, \citenamefont {Novario}, \citenamefont
  {Ringle},\ and\ \citenamefont {Schwarz}}]{Redshaw:2012jp}%
  \BibitemOpen
  \bibfield  {author} {\bibinfo {author} {\bibfnamefont {M.}~\bibnamefont
  {Redshaw}}, \bibinfo {author} {\bibfnamefont {G.}~\bibnamefont {Bollen}},
  \bibinfo {author} {\bibfnamefont {M.}~\bibnamefont {Brodeur}}, \bibinfo
  {author} {\bibfnamefont {S.}~\bibnamefont {Bustabad}}, \bibinfo {author}
  {\bibfnamefont {D.~L.}\ \bibnamefont {Lincoln}}, \bibinfo {author}
  {\bibfnamefont {S.~J.}\ \bibnamefont {Novario}}, \bibinfo {author}
  {\bibfnamefont {R.}~\bibnamefont {Ringle}}, \ and\ \bibinfo {author}
  {\bibfnamefont {S.}~\bibnamefont {Schwarz}},\ }\href {\doibase
  10.1103/PhysRevC.86.041306} {\bibfield  {journal} {\bibinfo  {journal} {Phys.
  Rev.}\ }\textbf {\bibinfo {volume} {C86}},\ \bibinfo {pages} {041306}
  (\bibinfo {year} {2012})}\BibitemShut {NoStop}%
\bibitem [{\citenamefont {Rahaman}\ \emph {et~al.}(2008)\citenamefont {Rahaman}
  \emph {et~al.}}]{Rahaman:2007ng}%
  \BibitemOpen
  \bibfield  {author} {\bibinfo {author} {\bibfnamefont {S.}~\bibnamefont
  {Rahaman}} \emph {et~al.},\ }\href {\doibase 10.1016/j.physletb.2008.02.047}
  {\bibfield  {journal} {\bibinfo  {journal} {Phys. Lett.}\ }\textbf {\bibinfo
  {volume} {B662}},\ \bibinfo {pages} {111} (\bibinfo {year} {2008})},\ \Eprint
  {http://arxiv.org/abs/0712.3337} {arXiv:0712.3337 [nucl-ex]} \BibitemShut
  {NoStop}%
\bibitem [{\citenamefont {Redshaw}\ \emph {et~al.}(2007)\citenamefont
  {Redshaw}, \citenamefont {Wingfield}, \citenamefont {McDaniel},\ and\
  \citenamefont {Myers}}]{Redshaw:2007un}%
  \BibitemOpen
  \bibfield  {author} {\bibinfo {author} {\bibfnamefont {M.}~\bibnamefont
  {Redshaw}}, \bibinfo {author} {\bibfnamefont {E.}~\bibnamefont {Wingfield}},
  \bibinfo {author} {\bibfnamefont {J.}~\bibnamefont {McDaniel}}, \ and\
  \bibinfo {author} {\bibfnamefont {E.~G.}\ \bibnamefont {Myers}},\ }\href
  {\doibase 10.1103/PhysRevLett.98.053003} {\bibfield  {journal} {\bibinfo
  {journal} {Phys. Rev. Lett.}\ }\textbf {\bibinfo {volume} {98}},\ \bibinfo
  {pages} {053003} (\bibinfo {year} {2007})}\BibitemShut {NoStop}%
\bibitem [{\citenamefont {Scielzo}\ \emph {et~al.}(2009)\citenamefont {Scielzo}
  \emph {et~al.}}]{Scielzo:2009nh}%
  \BibitemOpen
  \bibfield  {author} {\bibinfo {author} {\bibfnamefont {N.~D.}\ \bibnamefont
  {Scielzo}} \emph {et~al.},\ }\href {\doibase 10.1103/PhysRevC.80.025501}
  {\bibfield  {journal} {\bibinfo  {journal} {Phys. Rev.}\ }\textbf {\bibinfo
  {volume} {C80}},\ \bibinfo {pages} {025501} (\bibinfo {year} {2009})},\
  \Eprint {http://arxiv.org/abs/0902.2376} {arXiv:0902.2376 [nucl-ex]}
  \BibitemShut {NoStop}%
\bibitem [{\citenamefont {Rahaman}\ \emph {et~al.}(2011)\citenamefont
  {Rahaman}, \citenamefont {Elomaa}, \citenamefont {Eronen}, \citenamefont
  {Hakala}, \citenamefont {Jokinen}, \citenamefont {Kankainen}, \citenamefont
  {Rissanen}, \citenamefont {Suhonen}, \citenamefont {Weber},\ and\
  \citenamefont {Aysto}}]{Rahaman:2011zz}%
  \BibitemOpen
  \bibfield  {author} {\bibinfo {author} {\bibfnamefont {S.}~\bibnamefont
  {Rahaman}}, \bibinfo {author} {\bibfnamefont {V.~V.}\ \bibnamefont {Elomaa}},
  \bibinfo {author} {\bibfnamefont {T.}~\bibnamefont {Eronen}}, \bibinfo
  {author} {\bibfnamefont {J.}~\bibnamefont {Hakala}}, \bibinfo {author}
  {\bibfnamefont {A.}~\bibnamefont {Jokinen}}, \bibinfo {author} {\bibfnamefont
  {A.}~\bibnamefont {Kankainen}}, \bibinfo {author} {\bibfnamefont
  {J.}~\bibnamefont {Rissanen}}, \bibinfo {author} {\bibfnamefont
  {J.}~\bibnamefont {Suhonen}}, \bibinfo {author} {\bibfnamefont
  {C.}~\bibnamefont {Weber}}, \ and\ \bibinfo {author} {\bibfnamefont
  {J.}~\bibnamefont {Aysto}},\ }\href {\doibase 10.1016/j.physletb.2011.07.078}
  {\bibfield  {journal} {\bibinfo  {journal} {Phys. Lett.}\ }\textbf {\bibinfo
  {volume} {B703}},\ \bibinfo {pages} {412} (\bibinfo {year}
  {2011})}\BibitemShut {NoStop}%
\bibitem [{\citenamefont {Albert}\ \emph {et~al.}(2014)\citenamefont {Albert}
  \emph {et~al.}}]{Albert:2013gpz}%
  \BibitemOpen
  \bibfield  {author} {\bibinfo {author} {\bibfnamefont {J.~B.}\ \bibnamefont
  {Albert}} \emph {et~al.} (\bibinfo {collaboration} {EXO-200}),\ }\href
  {\doibase 10.1103/PhysRevC.89.015502} {\bibfield  {journal} {\bibinfo
  {journal} {Phys. Rev.}\ }\textbf {\bibinfo {volume} {C89}},\ \bibinfo {pages}
  {015502} (\bibinfo {year} {2014})},\ \Eprint {http://arxiv.org/abs/1306.6106}
  {arXiv:1306.6106 [nucl-ex]} \BibitemShut {NoStop}%
\bibitem [{\citenamefont {Menendez}(2018)}]{Menendez:2017fdf}%
  \BibitemOpen
  \bibfield  {author} {\bibinfo {author} {\bibfnamefont {J.}~\bibnamefont
  {Menendez}},\ }\href {\doibase 10.1088/1361-6471/aa9bd4} {\bibfield
  {journal} {\bibinfo  {journal} {J. Phys.}\ }\textbf {\bibinfo {volume}
  {G45}},\ \bibinfo {pages} {014003} (\bibinfo {year} {2018})},\ \Eprint
  {http://arxiv.org/abs/1804.02105} {arXiv:1804.02105 [nucl-th]} \BibitemShut
  {NoStop}%
\bibitem [{\citenamefont {Hyvarinen}\ and\ \citenamefont
  {Suhonen}(2015)}]{Hyvarinen:2015bda}%
  \BibitemOpen
  \bibfield  {author} {\bibinfo {author} {\bibfnamefont {J.}~\bibnamefont
  {Hyvarinen}}\ and\ \bibinfo {author} {\bibfnamefont {J.}~\bibnamefont
  {Suhonen}},\ }\href {\doibase 10.1103/PhysRevC.91.024613} {\bibfield
  {journal} {\bibinfo  {journal} {Phys. Rev.}\ }\textbf {\bibinfo {volume}
  {C91}},\ \bibinfo {pages} {024613} (\bibinfo {year} {2015})}\BibitemShut
  {NoStop}%
\bibitem [{\citenamefont {Carone}(1993)}]{Carone:1993jv}%
  \BibitemOpen
  \bibfield  {author} {\bibinfo {author} {\bibfnamefont {C.~D.}\ \bibnamefont
  {Carone}},\ }\href {\doibase 10.1016/0370-2693(93)90605-H} {\bibfield
  {journal} {\bibinfo  {journal} {Phys. Lett. B}\ }\textbf {\bibinfo {volume}
  {308}},\ \bibinfo {pages} {85} (\bibinfo {year} {1993})},\ \Eprint
  {http://arxiv.org/abs/hep-ph/9302290} {arXiv:hep-ph/9302290} \BibitemShut
  {NoStop}%
\bibitem [{\citenamefont {Bischer}\ \emph {et~al.}(2018)\citenamefont
  {Bischer}, \citenamefont {Rodejohann},\ and\ \citenamefont
  {Xu}}]{Bischer:2018zbd}%
  \BibitemOpen
  \bibfield  {author} {\bibinfo {author} {\bibfnamefont {I.}~\bibnamefont
  {Bischer}}, \bibinfo {author} {\bibfnamefont {W.}~\bibnamefont {Rodejohann}},
  \ and\ \bibinfo {author} {\bibfnamefont {X.-J.}\ \bibnamefont {Xu}},\ }\href
  {\doibase 10.1007/JHEP10(2018)096} {\bibfield  {journal} {\bibinfo  {journal}
  {JHEP}\ }\textbf {\bibinfo {volume} {10}},\ \bibinfo {pages} {096} (\bibinfo
  {year} {2018})},\ \Eprint {http://arxiv.org/abs/1807.08102} {arXiv:1807.08102
  [hep-ph]} \BibitemShut {NoStop}%
\bibitem [{\citenamefont {Rodejohann}\ and\ \citenamefont
  {Xu}(2019)}]{Rodejohann:2019quz}%
  \BibitemOpen
  \bibfield  {author} {\bibinfo {author} {\bibfnamefont {W.}~\bibnamefont
  {Rodejohann}}\ and\ \bibinfo {author} {\bibfnamefont {X.-J.}\ \bibnamefont
  {Xu}},\ }\href {\doibase 10.1007/JHEP11(2019)029} {\bibfield  {journal}
  {\bibinfo  {journal} {JHEP}\ }\textbf {\bibinfo {volume} {11}},\ \bibinfo
  {pages} {029} (\bibinfo {year} {2019})},\ \Eprint
  {http://arxiv.org/abs/1907.12478} {arXiv:1907.12478 [hep-ph]} \BibitemShut
  {NoStop}%
\bibitem [{\citenamefont {Arnold}\ \emph {et~al.}(2019)\citenamefont {Arnold}
  \emph {et~al.}}]{NEMO-3:2019gwo}%
  \BibitemOpen
  \bibfield  {author} {\bibinfo {author} {\bibfnamefont {R.}~\bibnamefont
  {Arnold}} \emph {et~al.} (\bibinfo {collaboration} {NEMO-3}),\ }\href
  {\doibase 10.1140/epjc/s10052-019-6948-4} {\bibfield  {journal} {\bibinfo
  {journal} {Eur. Phys. J.}\ }\textbf {\bibinfo {volume} {C79}},\ \bibinfo
  {pages} {440} (\bibinfo {year} {2019})},\ \Eprint
  {http://arxiv.org/abs/1903.08084} {arXiv:1903.08084 [nucl-ex]} \BibitemShut
  {NoStop}%
\bibitem [{\citenamefont {Arnold}\ \emph {et~al.}(2010)\citenamefont {Arnold}
  \emph {et~al.}}]{Arnold:2010tu}%
  \BibitemOpen
  \bibfield  {author} {\bibinfo {author} {\bibfnamefont {R.}~\bibnamefont
  {Arnold}} \emph {et~al.} (\bibinfo {collaboration} {SuperNEMO}),\ }\href
  {\doibase 10.1140/epjc/s10052-010-1481-5} {\bibfield  {journal} {\bibinfo
  {journal} {Eur. Phys. J.}\ }\textbf {\bibinfo {volume} {C70}},\ \bibinfo
  {pages} {927} (\bibinfo {year} {2010})},\ \Eprint
  {http://arxiv.org/abs/1005.1241} {arXiv:1005.1241 [hep-ex]} \BibitemShut
  {NoStop}%
\bibitem [{\citenamefont {Doi}\ \emph {et~al.}(1985)\citenamefont {Doi},
  \citenamefont {Kotani},\ and\ \citenamefont {Takasugi}}]{Doi:1985dx}%
  \BibitemOpen
  \bibfield  {author} {\bibinfo {author} {\bibfnamefont {M.}~\bibnamefont
  {Doi}}, \bibinfo {author} {\bibfnamefont {T.}~\bibnamefont {Kotani}}, \ and\
  \bibinfo {author} {\bibfnamefont {E.}~\bibnamefont {Takasugi}},\ }\href
  {\doibase 10.1143/PTPS.83.1} {\bibfield  {journal} {\bibinfo  {journal}
  {Prog. Theor. Phys. Suppl.}\ }\textbf {\bibinfo {volume} {83}},\ \bibinfo
  {pages} {1} (\bibinfo {year} {1985})}\BibitemShut {NoStop}%
\bibitem [{\citenamefont {Simkovic}\ \emph {et~al.}(1999)\citenamefont
  {Simkovic}, \citenamefont {Pantis}, \citenamefont {Vergados},\ and\
  \citenamefont {Faessler}}]{Simkovic:1999re}%
  \BibitemOpen
  \bibfield  {author} {\bibinfo {author} {\bibfnamefont {F.}~\bibnamefont
  {Simkovic}}, \bibinfo {author} {\bibfnamefont {G.}~\bibnamefont {Pantis}},
  \bibinfo {author} {\bibfnamefont {J.~D.}\ \bibnamefont {Vergados}}, \ and\
  \bibinfo {author} {\bibfnamefont {A.}~\bibnamefont {Faessler}},\ }\href
  {\doibase 10.1103/PhysRevC.60.055502} {\bibfield  {journal} {\bibinfo
  {journal} {Phys. Rev.}\ }\textbf {\bibinfo {volume} {C60}},\ \bibinfo {pages}
  {055502} (\bibinfo {year} {1999})},\ \Eprint
  {http://arxiv.org/abs/hep-ph/9905509} {arXiv:hep-ph/9905509 [hep-ph]}
  \BibitemShut {NoStop}%
\bibitem [{\citenamefont {Barea}\ \emph {et~al.}(2013)\citenamefont {Barea},
  \citenamefont {Kotila},\ and\ \citenamefont {Iachello}}]{Barea:2013bz}%
  \BibitemOpen
  \bibfield  {author} {\bibinfo {author} {\bibfnamefont {J.}~\bibnamefont
  {Barea}}, \bibinfo {author} {\bibfnamefont {J.}~\bibnamefont {Kotila}}, \
  and\ \bibinfo {author} {\bibfnamefont {F.}~\bibnamefont {Iachello}},\ }\href
  {\doibase 10.1103/PhysRevC.87.014315} {\bibfield  {journal} {\bibinfo
  {journal} {Phys. Rev.}\ }\textbf {\bibinfo {volume} {C87}},\ \bibinfo {pages}
  {014315} (\bibinfo {year} {2013})},\ \Eprint {http://arxiv.org/abs/1301.4203}
  {arXiv:1301.4203 [nucl-th]} \BibitemShut {NoStop}%
\bibitem [{\citenamefont {Suhonen}(2017)}]{Suhonen:2017krv}%
  \BibitemOpen
  \bibfield  {author} {\bibinfo {author} {\bibfnamefont {J.~T.}\ \bibnamefont
  {Suhonen}},\ }\href {\doibase 10.3389/fphy.2017.00055} {\bibfield  {journal}
  {\bibinfo  {journal} {Front. in Phys.}\ }\textbf {\bibinfo {volume} {5}},\
  \bibinfo {pages} {55} (\bibinfo {year} {2017})},\ \Eprint
  {http://arxiv.org/abs/1712.01565} {arXiv:1712.01565 [nucl-th]} \BibitemShut
  {NoStop}%
\bibitem [{\citenamefont {Haxton}\ and\ \citenamefont
  {Stephenson}(1984)}]{Haxton:1985am}%
  \BibitemOpen
  \bibfield  {author} {\bibinfo {author} {\bibfnamefont {W.~C.}\ \bibnamefont
  {Haxton}}\ and\ \bibinfo {author} {\bibfnamefont {G.~J.}\ \bibnamefont
  {Stephenson}},\ }\href {\doibase 10.1016/0146-6410(84)90006-1} {\bibfield
  {journal} {\bibinfo  {journal} {Prog. Part. Nucl. Phys.}\ }\textbf {\bibinfo
  {volume} {12}},\ \bibinfo {pages} {409} (\bibinfo {year} {1984})}\BibitemShut
  {NoStop}%
\end{thebibliography}%

\appendix

\section{Decay Rate of $\nu$SI-mediated double beta decay}
\label{sec:appA}
\noindent
Here we present the detailed derivation of the differential decay rate of \vSIbb{} decay,  pointing out the key differences and similarities in the context of standard $0\nu\beta\beta$ decay calculations.

\subsection{Leptonic Part}
\label{sec:appA-leptonic}
\noindent 
The leptonic part of $0\nu\beta\beta$ matrix element reads
\begin{align}
	i{M}_{0\nu}^{\mu\nu} &\approx \overline{\psi}_{e_2}\gamma^\mu P_L \frac{i}{\slashed{q}}(-i)m_{ee}\frac{i}{\slashed{q}}P_L\gamma^\nu\psi_{e_1^c},\nonumber\\
	&\approx
	\frac{im_{ee}}{q^{2}}\overline{\psi}_{e_2}\gamma^{\mu}\gamma^{\nu}P_{R}\psi_{e_1^{c}},
\label{eq:a-1}
\end{align}
where $m_{ee} = \sum U_{ei}^2 m_{\nu_i}$ is the usual effective mass with the neutrino masses $m_{\nu_i}$ and the charged-current leptonic mixing matrix $U$. The above expression is calculated using two massless neutrino propagators and one mass insertion. If the whole neutrino line in $0\nu\beta\beta$ decay was considered as a single propagator, then with the $P_L$ projectors one would obtain
\begin{align}
	\frac{i}{\slashed{q}}(-i)m_{ee}\frac{i}{\slashed{q}} \approx 
	\frac{i}{\slashed{q}-m_{ee}},
\end{align}
which is approximately equivalent for $|q^2| \gg m_{\nu_i}^2$.

By comparing the two diagrams in Fig.~\ref{fig:feyn} in the main text, the leptonic matrix element of \vSIbb{} decay can be written in an analogous way. Instead of a mass insertion one employs the $\nu$SI vertex, which additionally gives two extra external neutrino legs: 
%
\begin{align}
	i{M}_{\nu\text{SI}}^{\mu\nu} &\approx 
	\left[\overline{\psi}_{e_2}\gamma^\mu P_L \frac{i}{\slashed{q}}(-i)G_S\frac{i}{\slashed{q} - \slashed{p}}P_L\gamma^\nu\psi_{e_1^c}\right]
	\left[\overline{\psi}_{\nu_4}P_R\psi_{\nu_3^c}\right].
\end{align}
Here, $\psi_{\nu_3}$ and $\psi_{\nu_4}$ denote the external lines of neutrinos in the \vSIbb{} diagram and $p = p_{\nu_3} + p_{\nu_4}$ is the sum of final state neutrino momenta.  Note that we are here assuming a lepton number violating $\nu$SI interaction; the result for the conserving case is the same.
Assuming that the momenta of the final state leptons are negligible compared to the momentum $q$ of the neutrino propagators, $s = p^2 \ll q^2$, one can immediately relate the amplitude of \vSIbb{} decay to that of $0\nu\beta\beta$ decay,
\begin{equation}
	iM_{\nu{\rm SI}}^{\mu\nu} = iM_{0\nu}^{\mu\nu} 
	\frac{G_S}{m_{ee}}\overline{\psi}_{\nu_4} P_R \psi_{\nu_3^c}.
\label{eq:a}
\end{equation}
The above leptonic matrix elements are to be contracted with their nuclear counterparts. The structure of the latter is fully identical between the \vSIbb{} and $0\nu\beta\beta$ cases. With the same $q$-dependence giving rise to the same neutrino potential, the resulting NMEs are the same, $\mathcal{M}_{\nu\text{SI}} \approx \mathcal{M}_{0\nu}$. 

In calculating the leptonic phase space factor, we will take the $S_{1/2}$ approximation for the outgoing electrons and neutrinos,
\begin{align}
	\psi_e(p,\mathbf{x},s) &\approx 
	\sqrt{\frac{F_0(Z_f,p_e)}{(2\pi)^3 2E_e}} u_e(p,s)e^{i\vecl{p}\cdot\vecl{x}}, \\
	\psi_\nu(p,\mathbf{x},s) &\approx \frac{1}{\sqrt{(2\pi)^3 2E_\nu}} u_\nu(p,s)
	e^{i\vecl{p}\cdot\vecl{x}},
\label{eq:vplane} 
\end{align}
where $\vecl{p}$ and $E_{e,\nu}$ denote the electron and neutrino 3-momenta and energies, and $u_{e,\nu}$ stands for the usual Dirac spinor. For the electron wave function we include the Fermi function $F_0(Z_f,p_e)$, taking into account the interaction of the emitted electron with the final nucleus of charge $Z_f = Z+2$. It can be approximated for the purposes of our numerical calculations as \cite{Doi:1985dx}
\begin{equation}
\label{eq:fermi}
	F_0(Z_f, p_e) 
	= 4\frac{(2 p R)^{2(\gamma_0-1)}}{\left[\Gamma(1+2\gamma_0)\right]^2}
	  e^{\pi y}|\Gamma(\gamma_0+iy)|^2,
\end{equation}
with $\gamma_0 = \sqrt{1-(Z_f \alpha)^2}$ and $y = \alpha Z_f E_e / p$, where $\alpha$ denotes the fine-structure constant, $R \approx 1.2A^{1/3}$~fm is the nuclear radius ($A$ denotes the atomic number of the decaying isotope) and $\Gamma(x)$ is the Gamma function.

For a $0^{+}\rightarrow0^{+}$ nuclear transition and the $S_{1/2}$ approximation of the wave functions of the emitted electrons with momenta $p_1$ and $p_2$ we therefore have for the $0\nu\beta\beta$ matrix element
\begin{align}
	\left|i{M}_{0\nu}\right|^2 &=  |m_{ee}|^2 F^2(p_1,p_2) \nonumber\\
	&\times {\rm Tr}\left[\overline{u}_e P_R u_{e^c}\overline{u}_{e^c} P_L u_e\right] 
	\left|{\cal M}_{0\nu}\right|^2 \nonumber \\
 	&= |m_{ee}|^2 F^2(p_1,p_2) 2p_2\cdot p_1 
 	\left|{\cal M}_{0\nu}\right|^2.
\label{eq:x-19}
\end{align}
Here, ${\cal M}_{0\nu}$ denotes the nuclear part of the full matrix element and $F^2(p_1,p_2) = F_0(Z_f, p_1) F_0(Z_f, p_2)$. Consequently, combining Eq.~\eqref{eq:a} with the above leads to the \vSIbb{} matrix element
\begin{align}
	\left|i{M}_{\nu{\rm SI}}\right|^2 &= |G_S|^2 F^2(p_1,p_2) \nonumber\\
	&\times \left(2p_2\cdot p_1\right)\left(2p_3\cdot p_4\right) 
	\left|{\cal M}_{0\nu}\right|^2.
\label{eq:a-3}
\end{align}
In general, if the two outgoing neutrinos are replaced by any two massless fermions with a scalar product connected to the $G_S$ vertex, one would always get the product $(p_3\cdot p_4)$ in Eq.~\eqref{eq:a-3}.

With the above we can express the decay widths as
\begin{align}
	\Gamma_{0\nu} &= \left|\frac{m_{ee}}{m_e}\right|^2 \mathcal{G}_{0\nu} \left|\mathcal{M}_{0\nu}\right|^2, 
\label{eq:a-4} \\
	\Gamma_{\nu\text{SI}} &= \left|\frac{G_S m_e}{2R}\right|^2 \mathcal{G}_{\nu\text{SI}}
	\left|\mathcal{M}_{0\nu}\right|^2, 
\label{eq:a-5}
\end{align}
where the factor $m_e^2/(4R^2)$ is included to make the NME $\mathcal{M}_{0\nu}$ dimensionless and the phase space $\mathcal{G}_{\nu\text{SI}}$ have units of $\mathrm{yr}^{-1}$, and to conform to the usual conventions employed in $0\nu\beta\beta$ decay calculations. Furthermore, $\mathcal{G}_{0\nu}$ and $\mathcal{G}_{\nu\text{SI}}$ are the phase space factors, which we can derive starting with the following integrals 
\begin{align}
	I_{0\nu} &= \int F^2(p_1,p_2)(2p_1\cdot p_2) \nonumber\\
 	&\times \left[\prod_{i=1}^{2}\frac{d^3\vecl{p}_i}{(2\pi)^{3}2E_{i}}\right] 
 	\delta\left(E_I - E_F - \sum_{i=1}^2 E_i\right),
\label{eq:x-16} \\
	I_{\nu{\rm SI}} &= \int F^2(p_1,p_2) (2p_1\cdot p_2)(2p_3\cdot p_4) \nonumber\\
 	&\times \left[\prod_{i=1}^4\frac{d^3\vecl{p}_i}{(2\pi)^{3}2E_{i}}\right]
 	\delta\left(E_I - E_F - \sum_{i=1}^4 E_i\right),
\label{eq:x-17}
\end{align}
where $E_I$ and $E_F$ denote the energies of the initial and final nuclei, respectively, 
with the $Q$-value defined as $Q = E_I - E_F - 2m_e$.

Next, we transform the phase space integral from Cartesian to polar coordinates,
\begin{eqnarray}
 	&  & (2p_1 \cdot p_2)\left[\prod_{i=1}^{2}\frac{d^3\vecl{p}_i}{(2\pi)^3 2E_i}\right]
 	\nonumber\\
 	&=& 
 	\left[1-\frac{p_1 p_2 c_{21}}{E_1 E_2}\right]\frac{p_1^2 p_2^2 dp_1 dp_2 dc_{21}}{(2\pi)^4},
 \label{eq:s-8}
\end{eqnarray}
where in the second row we use $p_i$ to denote $|\vecl{p}_i|$ for simplicity, and $c_{ij}$ to denote the cosine of the angle between $\vecl{p}_i$ and $\vecl{p}_j$. With the above replacement and a similar one for $i = 3$ and $4$, the integrals in Eqs.~\eqref{eq:x-16} and \eqref{eq:x-17} become
\begin{eqnarray}
	I_{0\nu} & = & \int
	\delta\left(E_I - E_F - \sum_{i=1}^2 E_i\right) F^2(p_1,p_2) \nonumber\\
	&\times&
	\left[1-\frac{p_1 p_2 c_{21}}{E_1 E_2}\right]
	\frac{p_1^2 p_2^2 dp_1 dp_2 dc_{21}}{(2\pi)^4},
\label{eq:s-23}
\end{eqnarray}
and
\begin{eqnarray}
	I_{\nu{\rm SI}} & = & \int
	\delta\left(E_I - E_F - \sum_{i=1}^4 E_i\right) F^2(p_1,p_2) \nonumber \\
 	&\times&
 	\left[1-\frac{p_1 p_2 c_{21}}{E_1 E_2}\right]\frac{p_1^2 p_2^2 dp_1 dp_2 dc_{21}}{(2\pi)^4} \nonumber\\
	&\times&
	\left[1-\frac{p_3 p_4 c_{43}}{E_3 E_4}\right]\frac{p_3^2 p_4^2 dp_3 dp_4 dc_{43}}{(2\pi)^4}.
\label{eq:s-24}
\end{eqnarray}
Since the two final state neutrinos in \vSIbb{} decay are not visible, we need to integrate over their kinematic parameters $p_3$, $p_4$ and $c_{43}$. We include the $s$-channel dependence of $G_S$ given in Eq.~(\ref{eq:st-s}) in the main text and evaluate the following part of the phase space integral,
\begin{eqnarray}
	I_s &=& \int\left(\frac{1}{s - m_{\phi}^{2}}\right)^{2}\left(1 - \frac{p_3 p_4}{E_3 E_4}c_{43}\right) \nonumber\\
	&\times& \delta\!\left(E_{I}-E_{F}-\sum_{i=1}^{4}E_{i}\right)\!
	\frac{p_3^2 p_4^2 dp_3 dp_4 dc_{43}}{(2\pi)^4}.
\label{eq:s-1}
\end{eqnarray}
Assuming the two outgoing neutrinos are massless, we have $p_3 = E_3$, $p_4 = E_4$, $s = 2E_3 E_4(1 - c_{43})$, and thus
\begin{eqnarray}
	I_s &=& \int\frac{1-c_{43}}{(2E_3 E_4 (1-c_{43}) - m_\phi^2)^2} \nonumber \\
	 & &\times \delta\left(E_I - E_F - \sum_{i=1}^{4}E_i\right)  \nonumber\\
	 & &\times \frac{E_3^2 E_4^2 dE_3 dE_4 dc_{43}}{(2\pi)^4}.
\label{eq:s-2}
\end{eqnarray}
Integrating over $c_{43}$, $E_3$ and $E_4$ sequentially gives
\begin{equation}
	I_{s}(T_{12}) = \frac{Q - T_{12}}{4(2\pi)^4}\left(\xi\frac{2+\cos\xi}{\sin\xi}-3\right),
\label{eq:Is}
\end{equation}
with
\begin{equation}
	\xi = 2\arcsin\frac{Q - T_{12}}{m_\phi}.
\end{equation}
Eq.~\eqref{eq:s-24} then simplifies to
\begin{eqnarray}
	I_{\nu{\rm SI}}(T_{12}) &=& 
	\int m_\phi^4 \left(1-\frac{p_1 p_2 c_{21}}{E_1 E_2}\right) F^2(p_1,p_2) \nonumber \\
 	&  &\times\frac{p_1^2 p_2^2 dp_1 dp_2 dc_{21}}{(2\pi)^4} I_s(T_{12}).
\label{eq:s-4}
\end{eqnarray}
Note also that here $I_s(T_{12})$,  with $T_{12} = E_1 + E_2 - 2 m_e$, is an implicit function of $p_1$ and $p_2$.

In the limit of large mass $m_\phi$ of the assumed scalar mediator we have
\begin{equation}
	\lim_{m_\phi\to\infty} I_s(T_{12}) = \frac{(Q - T_{1/2})^5}{15m_\phi^4(2\pi)^4}.
\label{eq:s-11}
\end{equation}
Hence, collecting all the prefactors the resulting phase space factor reads
\begin{equation}
	\mathcal{G}_{\nu{\rm SI}} = \frac{2c_{\nu\text{SI}}}{15}\int\! dp_1 dp_2
	p_1^2p_2^2(Q-T_{12})^5 F^2(p_1,p_2),
\label{eq:GvSIap}
\end{equation}
with
\begin{equation}
	c_{\nu\text{SI}} = \frac{G_F^4\cos^4\theta_C}{256\pi^9 m_e^2}.
\end{equation}
The expression in Eq.~(\ref{eq:GvSIap}) is almost the same as the phase space factor of standard $2\nu\beta\beta$ decay when neglecting the final state lepton momenta in the corresponding nuclear matrix element, 
\begin{equation}
	\mathcal{G}_{\nu{\rm SI}} = \frac{1}{4\pi^2}\mathcal{G}_{2\nu}.
\end{equation}

For $0\nu\beta\beta$ decay, we can analogously write 
%
\begin{equation}
	\mathcal{G}_{0\nu} = c_{0\nu} \int dp_1 2p_1^2 p_2 E_2 F_0(E_{1})F_0(E_{2}),
\label{eq:0nuint}
\end{equation}
where $c_{0\nu}=G_F^4\cos^4\!\theta_C m_e^2/(16\pi^5)$. Note that employing the above definitions of the phase space factors and comparing Eq.~\eqref{eq:a-4} with Eq.~\eqref{eq:a-5} we get the ratio
\begin{equation}
	\frac{\Gamma_{\nu{\rm SI}}}{\Gamma_{0\nu}} \approx 
	\left|\frac{G_S m_e^2}{2 m_{ee} R}\right|^2
	\frac{\mathcal{G}_{\nu{\rm SI}}}{\mathcal{G}_{0\nu}},
\label{eq:ratio}
\end{equation}
relating the total decay widths of \vSIbb{} and $0\nu\beta\beta$ decay.

Using Eq.~\eqref{eq:s-4}, we also obtain the differential decay rate
\begin{eqnarray}
	\frac{d\Gamma_{\nu{\rm SI}}}{dp_1 dp_2 dc_{21}} &=& 
	c_{\nu\text{SI}}|m_\phi^2 G_S^0|^2
	\left|\mathcal{M}_{0\nu}\right|^2 
	\left(1-\frac{p_1 p_2c_{21}}{E_1 E_2}\right) \nonumber \\
	& &\times F^2(p_1,p_2) \frac{p_1^2 p_2^2 I_s(T_{12})}{(2\pi)^4}.
\label{eq:s-5}
\end{eqnarray}

If only the total kinetic energy of the electrons ($T = E_1 + E_2 - 2m_e$) is measured, as in most double beta decay experiments, one must calculate the corresponding differential rate $d\Gamma_{\nu{\rm SI}}/dT$ by integrating Eq.~\eqref{eq:s-5} over $dp_1$, $dp_2$ and $dc_{21}$ while keeping $T$ at a given value. Here the integral over $dc_{21}$ can be done analytically, while the $dp_1 dp_2$ part has to be evaluated numerically. 

Likewise, to derive the angular distribution $d\Gamma/dc_{12}$, we integrate Eq.~\eqref{eq:s-5} over $p_1$ and $p_2$ to  obtain the general form
\begin{align}
	\frac{d\Gamma}{dc_{12}} = \frac{\Gamma}{2}\left(1 - k_\theta c_{12}\right)
\end{align}
where $\Gamma$ is the total decay rate and $k_\theta$ the angular correlation for the mode in question (\vSIbb{}, $0\nu\beta\beta$, $2\nu\beta\beta$).
%

\subsection{Nuclear Part}
\label{sec:appA-nuclear}
\begin{table}
\centering
\vspace{0.3cm}
\begin{ruledtabular}
\begin{tabular}{ll}
NME & $\tilde{h}_\circ(q^2)$ \\
\hline
$\mathcal{M}_F = \langle h_{XX}(q^2) \rangle$		
& $\tilde{h}_{XX}(q^2) = \frac{1}{\left(1+q^2/m_V^2\right)^4}$  \tabularnewline
$\mathcal{M}^{\prime WW}_{GT} = 
\left\langle \frac{\vecl{q}^2}{m_p^2} h_{XX}(q^2) 
(\vecs{\sigma}_a\cdot\vecs{\sigma}_b) \right\rangle $		
& $\tilde{h}_{XX}(q^2)$                                         \tabularnewline
$\mathcal{M}^{\prime WW}_T = 
\left\langle \frac{\vecl{q}^2}{m_p^2} h_{XX}(q^2) \vecl{S}_{ab} \right\rangle$	
& $\tilde{h}_{XX}(q^2)$                                         \tabularnewline
$\mathcal{M}_{GT}^{AA} = \langle h_{AA}(q^2) (\vecs{\sigma}_a\cdot\vecs{\sigma}_b) \rangle$	
& $\tilde{h}_{AA}(q^2) = \frac{1}{\left(1+q^2/m_A^2\right)^4}$ \tabularnewline
$\mathcal{M}^{\prime AP}_{GT} =  
\left\langle \frac{\vecl{q}^2}{m_p^2} h_{AP}(q^2) 
(\vecs{\sigma}_a\cdot\vecs{\sigma}_b) \right\rangle$ 
& $\tilde{h}_{AP}(q^2) = \frac{(1 + q^2/m_\pi^2)^{-1}}{\left(1 + q^2/m_A^2\right)^4}$                                      \tabularnewline
$\mathcal{M}^{\prime AP}_T = 
\left\langle \frac{\vecl{q}^2}{m_p^2}  h_{AP}(q^2) \vecl{S}_{ab} \right\rangle$	
& $\tilde{h}_{AP}(q^2)$                                         \tabularnewline
$\mathcal{M}^{\prime\prime PP}_{GT} = \left\langle \frac{\vecl{q}^4}{m_p^4}
h_{PP}(q^2) (\vecs{\sigma}_a\cdot\vecs{\sigma}_b) \right\rangle$	
& $\tilde{h}_{PP}(q^2)= \frac{(1 + q^2/m_\pi^2)^{-2}}{\left(1+q^2/m_A^2\right)^4}$           \tabularnewline
$\mathcal{M}^{\prime\prime PP}_{T} = \left\langle \frac{\vecl{q}^4}{m_p^4}  
h_{PP}(q^2) \vecl{S}_{ab} \right\rangle$ 
& $\tilde{h}_{PP}(q^2)$                                         
\end{tabular}
\end{ruledtabular}
\caption{\label{tab:nmes} Definitions of the double beta decay Fermi ($\mathcal{M}_F$), Gamow-Teller ($\mathcal{M}_{GT}$) and tensor ($\mathcal{M}_T$) NMEs entering Eq.~(\ref{eq:s-16}) and the corresponding reduced form-factor product $\tilde{h}(q^2)$. We use here the usual notation $\langle \mathcal{O}_{ab} \rangle = \langle 0_F^+| \sum_{a \neq b} \tau^{+}_a\tau^{+}_b \mathcal{O}_{ab} |0_I^+\rangle$ with $\tau^+$ being the isospin-raising operator and $|0_I^+\rangle$ and $|0_F^+\rangle$ denoting the initial and final states of the nucleus, respectively. The $q$-dependent functions $h_\circ(q^2) = v(q^2)\tilde h_\circ(q^2)$ are enhanced by the neutrino potential of the standard light neutrino exchange in Eq.~(\ref{eq:neutrinopotentialLR}). The subscript $X$ collectively denotes the three possibilities $X = V, W, T_1$ sharing the same form factor shape parameter $m_V$. The Pauli matrices $\vecs{\sigma}_{a,b}$ stand for the spins of the individual nucleons $a$, $b$ and the tensor NMEs include the definition $\vecl{S}_{ab} = 3(\vecs{\sigma}_a \cdot \vecl{q}) (\vecs{\sigma}_b \cdot \vecl{q}) - (\vecs{\sigma}_a \cdot \vecs{\sigma}_b)$.}
\end{table}

\noindent
As illustrated above, under the very good approximation that the momenta of the final state leptons can be neglected compared to the momentum flow of the internal neutrino propagators, the resulting NMEs of \vSIbb{} and $0\nu\beta\beta$ decay will be identical. We here briefly summarize the method of calculation of the latter. For our numerical analysis we use the NME calculations in the IBM-2 \cite{Barea:2015kwa}, Shell Model \cite{Menendez:2017fdf} and QRPA \cite{Hyvarinen:2015bda} nuclear structure frameworks.

The key quantities entering the microscopic description of double beta decays are the nuclear matrix elements, values of which have to be obtained using demanding nuclear structure calculations. Let us identify now the elementary nuclear matrix elements necessary for computing  the exotic neutrinoless double beta decay mechanism introduced in this text.

Under the assumption of negligible momenta of the outgoing neutrinos the nuclear matrix element $|\mathcal{M}_{\nu\text{SI}}|$ entering Eq.~\eqref{eq:a-5} can be taken to be approximately equal to the nuclear matrix element of the standard mass mechanism, $\mathcal{M}_{\nu SI}\approx\mathcal{M}_{0\nu}$, thus allowing for writing the $\nu\text{SI}\beta\beta$ decay rate as in Eq.~\eqref{eq:gamma_vSI}. This approximation is reasonable, as the propagating momentum $p\sim10^{2}$~MeV, while the momenta of leptonic final states are of $\sim1$~MeV.

Following the standard literature \cite{Simkovic:1999re,Barea:2013bz} the NME $\mathcal{M}_{0\nu}$ for the $0^+\to 0^+$ transition can be written as
\begin{eqnarray}
\label{eq:nme_nu}
	\mathcal{M}_\nu &=& g_V^2 \mathcal{M}_F - g_A^2 \mathcal{M}_{GT}^{AA} 
	       + \frac{g_A g_{P}}{6} 
	         \left(\mathcal{M}^{\prime AP}_{GT} + \mathcal{M}^{\prime AP}_T\right) \nonumber\\
          &+& \frac{(g_V+g_W)^2}{12}
             \left(-2\mathcal{M}^{\prime WW}_{GT} + \mathcal{M}^{\prime WW}_T\right)  \nonumber\\
          &-& \frac{g_{P}^2}{48} 
             \left(\mathcal{M}^{\prime\prime PP}_{GT} 
                 + \mathcal{M}^{\prime\prime PP}_T
             \right).
             \label{eq:s-16}
\end{eqnarray}
In the above, $g_X = F_X(0)$ denotes the form factor charge, i.e.\ the value of the nuclear form factor $F_X(\vecl{q}^2)$ at zero momentum transfer. The standard values used for the charges read: $g_V=1$, $g_A=1.269$, $g_W=3.7$ and $g_P=231$. Especially crucial is the axial vector coupling $g_A$, quenching of which is the subject of ongoing discussions in the $0\nu\beta\beta$ community \cite{Suhonen:2017krv}. For the purpose of this work we consider the usually employed unquenched value for free nucleon, $g_A=1.269$.

The $q$-dependence arising from the product of the reduced form factors $F_X(q^2)/g_X$ is included in the nuclear matrix elements appearing in Eq.~(\ref{eq:nme_nu}). The individual Fermi ($\mathcal{M}_F$), Gamow-Teller ($\mathcal{M}_{GT}$) and tensor ($\mathcal{M}_T$) NMEs along with the associated reduced form factor products $\tilde h(q^2)$ are given in Tab.~\ref{tab:nmes}.

In addition to the product of the reduced nucleon form factors, the NMEs listed in Tab.~\ref{tab:nmes} also contain the so-called neutrino potential describing the $q$-dependence of the underlying particle physics mediator of $0\nu\beta\beta$ decay. In the standard formulation of \cite{Simkovic:1999re} and \cite{Barea:2013bz} the two-body transition operator is constructed in momentum space as the product of neutrino potential, $v(q)$, and the product of the reduced form factors $\tilde{h}(q^2)$. For the standard mass mechanism the neutrino potential reads
\begin{equation}
\label{eq:neutrinopotentialLR}
	v(q) = \frac{2}{\pi}\frac{1}{q(q + \tilde A)},
\end{equation}
where neutrino mass has been neglected in comparison with the typical internal neutrino momentum $q\sim 100$ MeV, and $\tilde{A}$ is the closure energy, which can be adopted from Ref.\  \cite{Haxton:1985am} or estimated using $\tilde{A}=1.12\,A^{1/2}$~MeV. The above expression therefore describes the long-range exchange of an essentially massless neutrino mediating the $0\nu\beta\beta$ decay.

\section{Interference with $2\nu\beta\beta$ Decay}
\label{sec:interference}
\noindent
Given the same initial and final states, the $\nu$SI-induced double beta decay will interfere with the SM $2\nu\beta\beta$ contribution. 
It will thus occur if two electron anti-neutrinos are emitted in \vSIbb{}.
As the phase space of \vSIbb{} in Eq.~\eqref{eq:s-24} differs from that of $2\nu\beta\beta$ only by an overall factor of $4\pi^2$, the phase space part of the interference contribution, $\mathcal{G}_{\nu{\rm SI}-2\nu}$ is given by $\mathcal{G}_{\nu{\rm SI}-2\nu} = 2\pi \mathcal{G}_{\nu{\rm SI}} = \frac{1}{2\pi} \mathcal{G}_{2\nu}$.

As for the nuclear part of the interference term, combining the amplitudes of \vSIbb{} and $2\nu\beta\beta$ decay gives a single power of the NME $\mathcal{M}_{0\nu}$ discussed above and a single power of the $2\nu\beta\beta$ NME, $\mathcal{M}_{2\nu}$.

Hence, the resulting decay rate for the interference contribution reads 
%
\begin{equation}
	\Gamma_{\nu{\rm SI}-2\nu} = \frac{G_S m_e}{2R} \mathcal{G}_{\nu{\rm SI}-2\nu} 2\mathrm{Re}[\mathcal{M}_{0\nu}^*\mathcal{M}_{2\nu}].
\end{equation}
Here we assume, as before, that the final state lepton momenta can be neglected in the calculation of the nuclear matrix elements of both \vSIbb{} and $2\nu\beta\beta$ decay. As discussed, this approximation is very good for \vSIbb{} as the nuclear scale $p_F$ is much larger than the $Q$-value. For $2\nu\beta\beta$ decay, the approximation is instead very rough as the only low-lying nuclear states are excited and the leptonic phase space and nuclear parts do not decouple. 
The $2\nu\beta\beta$ NME $\mathcal{M}_{2\nu}$ is dominated by the double Gamow-Teller transition and the corresponding values can be taken e.g.\ from Ref.~\cite{Barea:2015kwa}, where they were computed using the IBM-2 nuclear structure model.\\

\section{Details on the fit of NEMO-3 data}
\label{sec:nemo3-fitting}
\noindent
To perform the fit of NEMO-3 data, we extract the observed event
numbers from Fig.~3 of Ref.~\cite{NEMO-3:2019gwo} as well as the
theoretically expected numbers from Monte Carlo simulations. The
actual spectra shown in Ref.~\cite{NEMO-3:2019gwo} are different
from ours because they depend on various detector effects which cannot
be included without dedicated simulation of the detector. An
assumption we will take is that the relative size of event excess
or deficit is transferable to the theoretical distribution without detector effects.
For example, if the event number in a
bin in Fig.~3 of  Ref.~\cite{NEMO-3:2019gwo}  is 1\% higher than
the theoretical expectation, then we can assume that in our Fig.~\ref{fig:distortion} or \ref{fig:distortion-angular}
the measured value of $d\Gamma/dT$ or $d\Gamma/d\cos\theta$ is also
1\% higher than the blue curves. Based on this assumption, we can
convert the observed event numbers in Ref.~\cite{NEMO-3:2019gwo}.
Then by adding a small contribution of $\nu{\rm SI}\beta\beta$
to the $2\nu\beta\beta$ spectrum,
\begin{equation}
\frac{d\Gamma_{2\nu}}{dX}\rightarrow R_{X}\equiv\frac{d\Gamma_{2\nu}}{dX}+r^{2}\frac{d\Gamma_{\nu{\rm SI}}}{dX}, \label{eq:appb-2}
\end{equation}
where $X$ stands for either $T$ or $\cos\theta$ and $r$ is a small
number, we can perform a $\chi^{2}$-fit with respect to $r$. One
should note that even for the standard process, the total decay rate
has not been theoretically determined. Therefore  we adopt the following $\chi^{2}$-function to fit the data
\begin{equation}
\chi^{2}(r)=\sum_{i}\left(\frac{a\thinspace R_{X}^{i}-R_{X}^{i,{\rm obs}}}{\sigma_{i}}\right)^{2}, \label{eq:appb-3}
\end{equation}
where a scale factor $a$ has been introduced in front of  the theoretical
value $R_{X}^{i}$ ($i$ denotes the $i$-th bin) computed from Eq.~(\ref{eq:appb-2});
$R_{X}^{i,{\rm obs}}$ represents the observed values;
and $\sigma_{i}=R_{X}^{i,{\rm obs}}/\sqrt{N_{i}}$ with $N_{i}$ the
observed event number in the $i$-th bin.  In our $\chi^{2}$-fit,
when $r$ varies, we always use $a$ to rescale the distribution so
that the total rate $\sum_{i}a\thinspace R_{X}^{i}$ is a constant
equal to the standard 2$\nu\beta\beta$ total rate. The remaining
analysis is straightforward. 
Fitting to the data in Fig.~3 of Ref.~\cite{NEMO-3:2019gwo},
we find that $\Delta\chi^{2}(0.09)=1$, $\Delta\chi^{2}(0.13)=4$,
and $\Delta\chi^{2}(0.16)=9$, corresponding to 1, 2, and 3$\sigma$
limits, respectively. For the angular spectrum, we obtain $\Delta\chi^{2}(0.17)=1$,
$\Delta\chi^{2}(0.24)=4$, and $\Delta\chi^{2}(0.29)=9$, which implies
that the 1, 2, and 3$\sigma$ bounds are weaker than those from the
energy spectrum.
\end{document}